%% file: main.tex
\documentclass[journal]{IEEEtran}
\usepackage{amsmath,amsfonts}
\usepackage{algorithmic}
\usepackage{algorithm}
\usepackage{array}
\usepackage[caption=false,font=normalsize,labelfont=sf,textfont=sf]{subfig}
\usepackage{textcomp}
\usepackage{stfloats}
\usepackage{url}
\usepackage{verbatim}
\usepackage{graphicx}
\usepackage{cite}
\usepackage{orcidlink}

\usepackage{makecell, tabularx}
\usepackage{tabularx}
\usepackage{amsmath}
\usepackage{amsfonts}
\usepackage{tikz}
\usepackage{multirow}
\usepackage{xcolor}
\usepackage[export]{adjustbox}
\usepackage{hyperref}
\usepackage{subfig}
\usepackage[flushleft]{threeparttable}

\newcommand{\green}[1]{\textcolor[rgb]{0.0,0.5,0.0}{#1}}
\newcommand{\red}[1]{{\color{red}#1}}

\begin{document}

\title{Less is More: Skim Transformer \\ for Light Field Image Super-resolution}

\author{
Zeke Zexi Hu\orcidlink{0000-0003-4947-4832}, Haodong Chen \orcidlink{0000-0003-2254-5629}, Hui Ye \orcidlink{0009-0006-8882-1170}, Xiaoming Chen\orcidlink{0000-0002-7503-3021},~\IEEEmembership{Member, IEEE}, \\
Vera Yuk Ying Chung\orcidlink{0000-0002-3158-9650},~\IEEEmembership{Member, IEEE}, Yiran Shen\orcidlink{0000-0003-1385-1480},~\IEEEmembership{Senior Member, IEEE}, 
Weidong Cai\orcidlink{https://orcid.org/0000-0003-3706-8896},~\IEEEmembership{Member, IEEE}
\thanks{Zeke Zexi Hu, Haodong Chen, Hui Ye, Vera Yuk Ying Chung, and Weidong Cai are with the School of Computer Science, University of Sydney, Darlington, NSW 2008, Australia (E-mail: {zexi.hu, haodong.chen, huye0731, vera.chung, tom.cai}@sydney.edu.au).}
\thanks{Xiaoming Chen is with the School of Computer Science and Engineering, Beijing Technology and Business University, Beijing 102488, China (E-mail: xiaoming.chen@btbu.edu.cn).}
\thanks{Yiran Shen is with the School of Software, Shandong University, Jinan, 250100, China (E-mail: yiran.shen@sdu.edu.cn).}
\thanks{Corresponding Author: Xiaoming Chen.}
}

\markboth{Journal of \LaTeX\ Class Files,~Vol.~14, No.~8, August~2021}%
{Shell \MakeLowercase{\textit{et al.}}: A Sample Article Using IEEEtran.cls for IEEE Journals}


\newcommand{\LtFMamba}{L2FMamba}

\newif\ifhl
% \hltrue % comment this out for the camera-ready version

\ifhl
    % Highlighting enabled
    \newcommand{\hl}[1]{\textcolor{blue}{#1}}  
    \newenvironment{highlight}{\color{blue}}{}
\else
    % Highlighting disabled
    \newcommand{\hl}[1]{#1}
    \newenvironment{highlight}{}{}
\fi

\maketitle

\input{tex/0.Abstract.tex}

\begin{IEEEkeywords}
Light field, Super-resolution, Image processing, Deep learning.
\end{IEEEkeywords}

\input{tex/1.Introduction.tex}
\input{tex/2.RelatedWorks.tex}
\input{tex/3.Methodology.tex}
\input{tex/4.Experiments.tex}
\input{tex/5.Conclusion.tex}

\section{Acknowledgment}
This work was supported in part by the Natural Science Foundation of China (No. 62577004) and the Open Project Program of State Key Laboratory of Virtual Reality Technology and Systems, Beihang University (VRLAB2025C04).

\bibliographystyle{IEEEtran}
\bibliography{reference}

\input{tex/Bio}

\end{document}

%% file: tex/0.Abstract.tex
\begin{abstract}
A light field image captures scenes through its micro-lens array, providing a rich representation that encompasses spatial and angular information. While this richness comes at significant data redundancy, most existing methods tend to indiscriminately utilize all the information from sub-aperture images (SAIs) in an attempt to harness every visual cue regardless of their disparity significance. However, this paradigm inevitably leads to disparity entanglement, a fundamental cause of inefficiency in light field image processing. To address this limitation, we introduce the Skim Transformer, a novel architecture inspired by the ``less is more" philosophy. It features a multi-branch structure where each branch is dedicated to a specific disparity range by constructing its attention score matrix over a skimmed subset of SAIs, rather than all of them. Building upon it, we present SkimLFSR, an efficient yet powerful network for light field image super-resolution. Requiring only 67\% of \hl{the prior leading method's parameters}, SkimLFSR achieves state-of-the-art results surpassing the best existing method by 0.63 dB and 0.35 dB PSNR at the $2\times$ and $4\times$ tasks, respectively. Through in-depth analyses, we reveal that SkimLFSR, guided by the predefined skimmed SAI sets as prior knowledge, demonstrates distinct disparity-aware behaviors in attending to visual cues. \hl{Last but not least, we conduct an experiment to validate SkimLFSR's generalizability across different angular resolutions, where it achieves competitive performance on a larger angular resolution without any retraining or major network modifications.} These findings highlight its effectiveness and adaptability as a promising paradigm for light field image processing.
\end{abstract}

%% file: tex/1.Introduction.tex
\section{Introduction} \label{sec:Introduction}
Light Field (LF) imaging captures light rays of a scene from multiple angular directions in a single shot, enriching the scene representation and enabling advanced computer vision capabilities such as post-capture refocusing \cite{fiss2014refocusing}, depth estimation \cite{heberUshapeICCV2017, wangOcclusionawareDepthEstimation2015}, material recognition \cite{wangLFRecognition_ECCV2016, luLFRecognition_2019}, salient object detection \cite{chen_TMM2023}, microscopy \cite{verinaz_TCI2023, levoy2006light}, and virtual reality (VR) media enhancement \cite{broxton2020immersive}. The development of LF acquisition devices spans from early laboratory-built camera arrays \cite{GoogleLF, vaishSTFgantry_2008} to practical cameras, such as Raytrix \cite{Raytrix} and Lytro Illum \cite{Lytro}, and the more recent panoramic LF system \cite{broxton2020immersive} built by Google.

Due to technical limitations, accommodating a large number of micro-lenses within a compact form often comes at the cost of spatial resolution in LF images. For instance, the spatial resolution of a Lytro Illum camera is down to $376 \times 541$. To mitigate this limitation, several light field image super-resolution (LFSR) methods have been proposed to enhance the spatial resolution while preserving LF's intrinsic angular structure. Inherently, the LF's angular structure is non-uniform and can vary significantly across different spatial locations and angular directions. These variations arise primarily from two factors: scene depth (from an image-wise perspective) and camera configuration (from a device perspective). Disparity modeling plays a crucial role in harnessing this angular information to achieve effective LFSR.

Fig. \ref{fig:disparity} exemplifies how parallax manifests differently across disparity ranges. Specifically, Fig. \ref{fig:disparity} (b) demonstrates a larger disparity range by computing the parallax between two outer sub-aperture images (SAIs), making foreground objects such as the Lego studs (the blue box) appear more visually prominent. In contrast, Fig. \ref{fig:disparity} (c), which reflects a smaller disparity range using inner SAIs, renders these structures more blurred and indistinct. Conversely, background elements like the brick wall edges (the red box) exhibit excessive parallax in Fig. \ref{fig:disparity} (b), whereas they appear sharper and more coherent in Fig. \ref{fig:disparity} (c). These disparities reveal nuanced variations and complexity of disparity distribution in LF images, meanwhile, offering valuable cues for enhancing LFSR performance.

\input{tex/figures/disparity.tex}

The recent integration of deep learning techniques, especially Vision Transformers (ViT) techniques \cite{liangEPIT_ICCV2023, congLFDET_TMM2024, liangLFT_SPL2022, hu_M2MTNet_TMM2024}, has demonstrated promising potential in leveraging the self-attention mechanism to establish long-range dependencies for spatial, angular, and epipolar plane image (EPI) subspaces in LF images.
However, existing Transformer-based methods commonly overlook the substantial disparity variations inherent in LF data. These methods typically apply self-attention uniformly across the entire LF tensor in a single pass, attempting to exploit all available cues without adequately considering their significance, instead relying on the self-attention mechanism to learn the relevance solely from training data.

In this paper, we identify this issue as disparity entanglement, where heterogeneous disparity cues are processed homogeneously. This not only impairs a LFSR network’s ability to model disparity but also introduces computational redundancy. To address this issue, we propose the Skim Transformer, a novel Transformer architecture grounded in the \textbf{``less is more"} philosophy. Rather than indiscriminately processing all SAIs, the Skim Transformer selectively samples a subset of SAIs, referred to as a skimmed SAI set, that serves as prior knowledge to effectively manage disparity information and guide the construction of self-attention. Meanwhile, a multi-branch structure is constructed to broaden this capability to handle diverse disparity ranges.
As a result, the Skim Transformer processes less information, but not only mitigates computation redundancy but also disentangles disparity to extract more relevant information.

Building upon Skim Transformers, we present SkimLFSR, an efficient and effective LFSR network that not only delivers superior performance but also improves efficiency through its paradigm of disparity disentanglement. In our experiments, a lightweight variant of SkimLFSR outperforms nearly all existing methods \cite{liangEPIT_ICCV2023,congLFDET_TMM2024} while using only 37\% of the parameters, 35\% of the FLOPs, and 28\% of the inference time. The full version of SkimLFSR surpasses the previous state-of-the-art \cite{hu_M2MTNet_TMM2024} by a substantial margin of 0.63 dB and 0.35 dB in PSNR for the $2 \times$ and $4 \times$ LFSR tasks, respectively, while requiring just 67\% of the parameters used by the prior leading method.

We further conduct a series of in-depth analyses to uncover the inner workings of SkimLFSR. The results reveal that SkimLFSR's superior performance originates from its disparity-aware characteristic, as evidenced through deep feature analyses where SkimLFSR exhibits latent discriminative capability toward scene depths and camera configurations. Remarkably, this capability emerges even though neither of these two modalities is explicitly provided as supervision during the regression-based LFSR training, highlighting the effectiveness of Skim Transformer to implicitly capture and exploit underlying disparity cues.

\hl{Finally, as Skim Transformers projects only a skimmed SAI set instead of the entire angular space into embedding, it is decoupled from the angular resolution that it is trained on, and as a result, it is inherently angular-resolution-agnostic, i.e., it can generalize to different angular resolutions. In the experiments, we will validate that Skim Transformers learn essential disparity information that is generalizable across different angular resolutions, which enables SkimLFSR to achieve competitive performance on an angular resolution that is different from the one it is trained on without any retraining and major network modifications.}

In summary, the contributions of this work are as follows:
\begin{enumerate}
    \item We identify the issue of disparity entanglement in existing Transformer-based LFSR methods, where heterogeneous disparity cues are processed homogeneously, leading to suboptimal performance and computational redundancy.
    \item We propose the Skim Transformer, a novel architecture built on the ``less is more" philosophy. By selectively sampling a skimmed SAI set in self-attention and adopting a multi-branch structure, Skim Transformer explicitly targets different disparity ranges and achieves disparity disentanglement.
    \item Built upon this, we present SkimLFSR, an efficient and high-performing LFSR network. It surpasses the previous leading method, while requiring substantially fewer parameters and shorter inference time, striking a good balance between performance and efficiency.
    \item Through in-depth analyses, we reveal that SkimLFSR's superior performance stems from its disparity-aware characteristics and latent discriminative capability toward scene depth and camera configuration, learned implicitly from a regression-based task.
    \item \hl{We further discover that the Skim Transformer is inherently angular-resolution-agnostic, allowing SkimLFSR to seamlessly generalize to different angular resolutions and achieve very competitive performance without any retraining and major network modifications.}
\end{enumerate}

%% file: tex/figures/disparity.tex
\begin{figure}
\begin{center}
    \tabcolsep=0.04cm
    \renewcommand{\arraystretch}{0.6}
    \begin{tabular}{lrlrlr}
    \multicolumn{2}{c}{\scriptsize (a) Raw image} &
    \multicolumn{2}{c}{\scriptsize (b) Large disparity range} &
    \multicolumn{2}{c}{\scriptsize (c) Small disparity range} \\
    \multicolumn{2}{c}{\includegraphics[width=0.158\textwidth, height=0.130\textwidth]{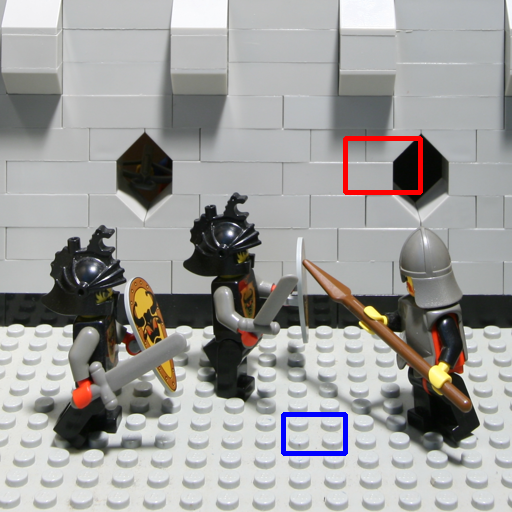}} &
    \multicolumn{2}{c}{
        \begin{tikzpicture}
            \node[anchor=south west,inner sep=0] (background) at (0,0) {
                \includegraphics[width=0.158\textwidth, height=0.130\textwidth]{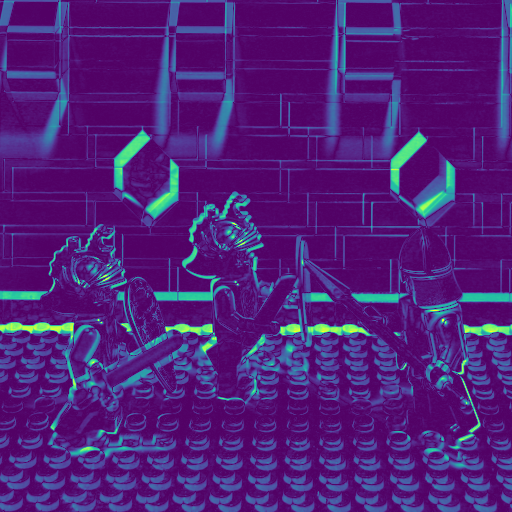}
            };
            \begin{scope}[x={(background.south east)},y={(background.north west)}]
                \node[anchor=south east,inner sep=0] at (1,0) {
                    \includegraphics[width=0.040\textwidth]{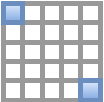}
                };
            \end{scope}
        \end{tikzpicture}
    } &
    \multicolumn{2}{c}{
        \begin{tikzpicture}
            \node[anchor=south west,inner sep=0] (background) at (0,0) {
                \includegraphics[width=0.158\textwidth, height=0.130\textwidth]{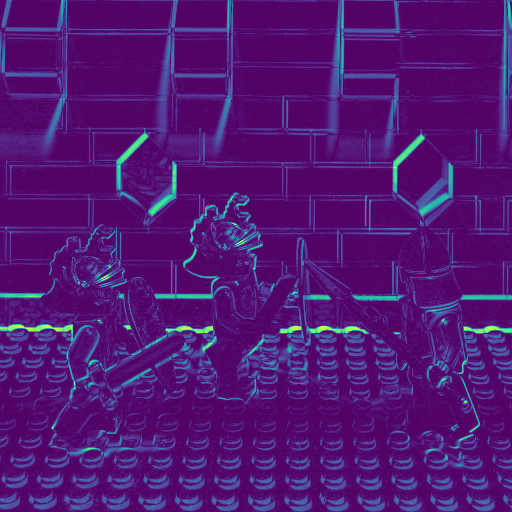}
            };
            \begin{scope}[x={(background.south east)},y={(background.north west)}]
                \node[anchor=south east,inner sep=0] at (1,0) {
                    \includegraphics[width=0.040\textwidth]{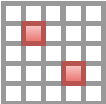}
                };
            \end{scope}
        \end{tikzpicture}
    } \\

    \includegraphics[width=0.073\textwidth, height=0.045\textwidth,cfbox=blue 1pt 0pt]{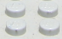} &
    \includegraphics[width=0.073\textwidth, height=0.045\textwidth,cfbox=red 1pt 0pt]{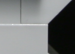} &

    \includegraphics[width=0.073\textwidth, height=0.045\textwidth,cfbox=blue 1pt 0pt]{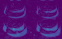} &
    \includegraphics[width=0.073\textwidth, height=0.045\textwidth,cfbox=red 1pt 0pt]{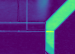} &

    \includegraphics[width=0.073\textwidth, height=0.045\textwidth,cfbox=blue 1pt 0pt]{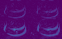} & 
    \includegraphics[width=0.073\textwidth, height=0.045\textwidth,cfbox=red 1pt 0pt]{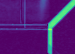} \\
    \vspace{-0.3cm}
    \\
    \end{tabular}
    \caption{Parallax at different disparity ranges in \textit{Lego Knights}. The bottom-right indicator in (b) and (c) indicates the SAIs for calculating the parallax.}
    \vspace{-15pt}
    \label{fig:disparity}
\end{center}
\end{figure}

%% file: tex/2.RelatedWorks.tex
\section{Related Work} \label{sec:RelatedWorks}
Processing 4D LF images introduces substantial challenges for neural network design due to their high dimensionality and the intricate spatial-angular dependencies they encode. To alleviate these issues, Wang et al. \cite{wangLFRecognition_ECCV2016} introduced the interleaved filter for LF material recognition, which approximates 4D convolutions by decomposing them into separate spatial and angular convolutions.

This decomposition paradigm was subsequently adapted for LFSR by Yoon et al. \cite{yoon2017LFCNN} in their LFCNN consisting of two separate sub-networks. Yeung et al. \cite{yeungSAS_LFSR2019} further improve this paradigm by making the network end-to-end. Building on these foundations, Jin et al. \cite{jinLFSSRATO_2020} introduced an all-to-one strategy, where each SAI is super-resolved using contextual information from all other SAIs, with a structure-aware loss to preserve parallax. Wang et al. \cite{wangLfInterNet_ECCV2020} proposed a two-branch network that separately extracts spatial and angular features and fuses them iteratively. Liu et al. \cite{liuLFIINet_TMM2021} further improved angular-spatial modeling by incorporating dilated convolutions in a multi-scale pyramid structure to expand receptive fields across spatial and angular subspaces. Chen et al. \cite{chen_TMM2021} explored an approach incorporating frequency-domain learning and semantic priors.

Given that LF images inherently encompass multiple subspaces, the decomposition paradigm is extended beyond the spatial and angular subspaces. Cheng et al. \cite{chengLFSSRSAV_TCI2022} proposed LFSSR-SAV, which employs spatial-angular versatile convolution to handle EPI subspaces. Hu et al. \cite{huDKNet_TIM2022} generalized this idea with the Decomposition Kernel Network (DKNet), jointly modeling spatial, angular, and EPI subspaces. Wang et al. \cite{wangDistgSSR_TPAMI2023} introduced a disentanglement mechanism to better aggregate and refine multi-subspace features, while Duong et al. \cite{duongHLFSR_TCI2023} proposed a hybrid model that integrates multi-orientation EPI extractors with angular and spatial representation extractors. \hl{Zhang et al. \cite{zhangCascadeResidualLearning_2025} proposed a cascade residual learning based adaptive feature aggregation framework to aggregate features from the subspaces. This paradigm is also adopted in other LF-related tasks. Yeung et al. \cite{yeungSAS_ECCV2018} proposed a spatial-angular separable convolution, consisting of a spatial convolution and an angular convolution, for LF angular super-resolution. In contrast to this task, typically performed on a single LF image with a limited field of view (FOV), Mao et al. \cite{mao2025deep} introduced a novel LF task called sparse-to-dense inbetweening, which aims to reconstruct a high–angular-resolution LF image from two low–angular-resolution inputs to extend the FOV. In their proposed method, the feature extraction process is decomposed into two EPI blocks and one spatial block.}

\hl{Parallel to advancements in SISR, several techniques have been extended to LFSR, such as degradation estimation \cite{bell2019blind, zhangBlindImageSuperResolution_2021}, which has been incorporated into LFSR pipelines by Wang et al. \cite{wang2024real} and Xiao et al. \cite{xiao2025incorporating} to improve robustness against diverse degradation types. CutBlur \cite{yoo_CutBlur_CVPR2020} is a data augmentation strategy that copies patches from LR images and pastes them into HR images, encouraging the model to learn which regions require super-resolution rather than uniformly enhancing all pixels. Inspired by CutBlur, Xiao et al. \cite{xiaoCutMIB_CVPR2023} introduced a “cutting–blending–pasting” strategy that blends LR patches across different SAIs to enable the same capability in LFSR methods.}

The ViT architecture \cite{ViT, liangSwinIR_ICCV2021, chenHAT_CVPR2023, zhouSRFormer_ICCV2023} has been a popular architecture for SISR and has expanded its successes to LFSR. Wang et al. \cite{wangDPT_AAAI2022} proposed DPT equipped with content and gradient Transformers to build long-range dependencies within the spatial subspace. Liang et al. \cite{liangLFT_SPL2022} employed self-attention on overlapped local windows in their spatial Transformers, akin to SwinIR \cite{liangSwinIR_ICCV2021} and HAT \cite{chenHAT_CVPR2023} for SISR. Transformers are also extended to the EPI subspaces in EPIT \cite{liangEPIT_ICCV2023}. Cong et al. proposed LF-DET \cite{congLFDET_TMM2024}, integrating a sub-sampling spatial strategy and a multi-scale angular modeling strategy to enhance spatial and angular modeling. Hu et al. \cite{hu_M2MTNet_TMM2024} proposed M2MTNet with Many-to-Many Transformers that avoid the practice of merging subspaces into the batch dimension and resolve the broadly existing issue of subspace isolation. \hl{Xiao et al. \cite{xiao_OHT_AAAI2025} incorporated SAI-wise occlusion maps as an additional branch and fused them with the initial visual features to achieve occlusion-aware LFSR.}

\hl{More recently, Mamba-based methods \cite{gu_Mamba_arxiv2023, dao_Mamba2_arxiv2024} have emerged as a promising alternative to Transformer-based methods. Its Selective State Space Models (SSM) uses input-dependent selective scanning to propagate a latent state across the sequence, and its computational complexity is linear with the sequence length, compared to the quadratic complexity of Transformer-based methods. This property has attracted the computer vision community to adopt it in various tasks \cite{zhu_VisionMamba_ICML2024, guo_MambaIR_ECCV2024}. For LFSR, Xia et al. \cite{xia_LFMamba_arxiv2024} proposed LFMamba, which adopts SSM to model the spatial, angular, and EPI subspaces as sequences. While the EPI subspace is for modeling spatial-angular correlations, Wei et al. \cite{wei_L2FMamba_TCI2025} argue that EPI-based methods are limited by locality constraints because their information comes from a single epipolar line. They proposed \LtFMamba, which adopts SSM to directly perform on individual SAIs, across SAIs, and on the macropixel image (MacPI). Liu et al. \cite{liu2025l3fmamba} proposed L3FMamba, which adopts a SSM-based method with dark, bright, and average channel prior injection for low-light LF image enhancement.}

Despite these recent advancements, as discussed in the previous section, a key limitation remains: most existing methods treat all the disparity cues indiscriminately, leading to disparity entanglement. Our Skim Transformer will be introduced to address this issue.

%% file: tex/3.Methodology.tex
\input{tex/figures/network.tex}

\section{Methodology} \label{sec:Methodology}
\subsection{Preliminaries}
The LFSR task aims to enhance the spatial resolution of a low-resolution (LR) LF image $I_{LR}$ and produce a super-resolved (SR) LF image $I_{SR}$ to approximate the corresponding high-resolution (HR) LF image $I_{HR}$:
\begin{equation}
\begin{split}
 I_{SR} = \mathcal{F}(I_{LR}), & \quad I_{LR} \in \mathbb{R}^{U \times V \times W \times H \times C}, \\
       & \quad I_{SR} \in \mathbb{R}^{U \times V \times rW \times rH \times C}
\end{split}
\end{equation}
where $\mathcal{F}(\cdot)$ is the LFSR function, $(U \times V)$ and $(W \times H)$ represent the angular and spatial subspaces of a LR image, respectively. The term $r$ denotes the LFSR scale.

\subsection{Network Architecture}
The proposed SkimLFSR network is illustrated in Fig. \ref{fig:Network}. The network comprises three primary stages: initial feature extraction, deep feature extraction, and image generation. The initial feature extraction stage utilizes four consecutive $3 \times 3$ convolution layers operating on the spatial subspace of LF images to obtain low-level features. The subsequent deep feature extraction stage incorporates a sequence of $N_{CB}$ correlation blocks to derive comprehensive correlation information and establish a high-level spatial-angular representation. The final image generation stage aggregates the deep features via convolution layers and upsamples the spatial resolution through a pixel shuffler \cite{shiESPCN_CVPR2016} to produce the final LF image.

\noindent\textbf{Correlation Blocks.} The correlation block comprises two specialized Transformers: the Skim Transformer, which operates in the spatial subspace, and the angular Transformer, which focuses on the angular subspace. The angular Transformer follows the methodologies of prior works \cite{liangLFT_SPL2022, congLFDET_TMM2024} at large, utilizing a vanilla Transformer to build long-range dependencies in the angular subspace with each SAI as a token.

\noindent\textbf{Connection Enhancement.} To improve SkimLFSR's data flow, besides a standard skip connection that bridges the SR and LR, two extra connections are integrated, as illustrated in Fig. \ref{fig:Network}. Firstly, in alignment with DKNet \cite{huDKNet_TIM2022}, a raw image connection concatenates the raw image tensor directly to the output of the last correlation block prior to the image generation stage. This connection acts as a special form of dense connections \cite{heResNet_CVPR2016}, ensuring that raw image data directly contributes to feature aggregation and upsampling. Secondly, inspired by RGT \cite{chenRGT_ICLR2024}, each correlation block incorporates a learnable skip connection that dynamically adjusts the input by channel-wise coefficients $\alpha \in \mathbb{R} ^ C$. These two connections incur negligible computational overhead but empirically improve the network performance by about 0.1 dB PSNR.

\subsection{Skim Transformer}
\input{tex/figures/SkimTransformer.tex}

In this subsection, we present the core component of SkimLFSR, Skim Transformers. An illustration is presented in Fig. \ref{fig:SkimTransformers}. It features a multi-branch structure to target different disparity ranges.

Given a 4D LF tensor $\mathbf{X} \in \mathbb{R}^{U \times V \times W \times H \times C}$, the channels are at first split evenly into $N_{DSA}$ branches, with each branch $X_i$ holding $C / N_{DSA}$ channels. Each $X_i \in \mathbb{R}^{U \times V \times W \times H \times C / N_{DSA}}$ will undergo Disparity Self-attention $DSA_i$ to explicitly construct long-range dependencies for a specific disparity range.

\noindent\textbf{Matrix Projection.} To achieve this, diverging from conventional LF Transformers \cite{congLFDET_TMM2024,liangLFT_SPL2022,liangEPIT_ICCV2023}, we introduce two substantial revamps for the matrix projection of query, key, and value matrices:

First, the query and key matrices, denoted as $\mathcal{Q}_i$ and $\mathcal{K}_i$, are constructed on a skimmed SAI set $\bar{X}_i$ holding $\mathcal{S}_i$ SAIs out of the full SAI set $\{U \times V\}$, i.e. $\{\mathcal{S}_i\} \subseteq \{U \times V\}$. This skimming process will serve as a selective sampling of the angular subspace to target specific disparity ranges.

Second, to facilitate a 4D LF tensor to be processed by the 1D self-attention mechanism, conventional LF Transformers typically merge the angular subspace with the batch dimension, treating the spatial subspace as tokens and the channel dimension as embedding, i.e. $\bar{X}_i \in \mathbb{R}^{UV \times WH \times C'}$ $(C'=C/N_{DSA})$. In contrast, the proposed Disparity Self-attention $DSA_i$ merges the angular subspace with the channel dimension, resulting in a LF tensor $\bar{X}_i \in \mathbb{R}^{WH \times \mathcal{S}_i C'}$, with the spatial subspace remained as tokens, the merged channel dimension representing the disparity information, and the batch dimension omitted. This design enables the subsequent disparity embedding process to encode disparities implicitly.

\noindent\textbf{Disparity Embedding.} Disparity embedding is accomplished via a linear layer $D_i$ projecting from $\mathbb{R}^{\mathcal{S}_i C / N_{DSA}}$ to $\mathbb{R}^{C_D}$ to produce a compact disparity embedding $\mathcal{D}_i \in \mathbb{R}^{WH \times C_D}$. This embedding is then fed to two linear layers, $Q_i$ and $K_i$, each of which projects from $\mathbb{R}^{C_D}$ to $\mathbb{R}^{C_{QK}}$, to generate the query and key matrices $\mathcal{Q}_i$ and $\mathcal{K}_i$, respectively. On the other hand, the value matrix, $\mathcal{V}_i$, is directly derived from $X_i$, thus preserving the entire angular information with the full SAI set.

\noindent\textbf{Self-attention Output.} As a result, the process for obtaining these three matrices for self-attention is formally expressed as follows:
\begin{gather}
    \mathcal{Q}_i = Q_i (\mathcal{D}_i), \; \mathcal{K}_i = K_i (\mathcal{D}_i), \; \mathcal{V}_i = X_i, \; \\
    \mathcal{D}_i = D_i (\bar{X}_i); \\
 X_i \in \mathbb{R}^{U \times V \times W \times H \times C / N_{DSA}}; \; \\
    \quad \bar{X}_i \in \mathbb{R}^{WH \times \mathcal{S}_i C / N_{DSA}}, \; \{\mathcal{S}_i\} \subseteq \{U \times V\}; \\
 Q_i, K_i: \mathbb{R}^{C_D} \mapsto \mathbb{R}^{C_{QK}}, \label{eq:qk} \\
 D_i: \mathbb{R}^{\mathcal{S}_i C / N_{DSA}} \mapsto \mathbb{R}^{C_D}. \label{eq:disparity_embedding}
\end{gather}

With the obtained query, key, and value matrices, the upcoming self-attention calculation, including the attention score matrix and the self-attention output, follows the typical Transformer method:
\begin{equation}
 DSA_i(\mathcal{Q}_i,\mathcal{K}_i,\mathcal{V}_i) = Softmax(\mathcal{Q}_i \mathcal{K}_i^T / \sqrt{C_{QK}}) \mathcal{V}_i,
\end{equation}
where $DSA_i \in \mathbb{R}^{WH \times UVC / N_{DSA}}$. Note that as the channel number is made consistent as $C/N_{DSA}$ through the Skim Transformer, the shape of self-attention output aligns with the input. Therefore, channel projection is not necessary, and the feed-forward network can be removed to reduce the computation.

Finally, DSA's output is compiled by concatenation as the final output of the Skim Transformer ($SkimT$):
\begin{equation}
    \mathbf{Y} = SkimT(\mathbf{X}) = [DSA_1, DSA_2, \cdots, DSA_{N_{DSA}}]
\end{equation}
where $[\cdot]$ signifies the concatenation operation, and $\mathbf{Y} \in \mathbb{R}^{WH \times UVC}$ and finally $\mathbb{R}^{U \times V \times W \times H \times C}$.

\input{tex/tables/computation.tex}

\subsection{Discussion} \label{sec:Discussion}
With these innovations, the Skim Transformer confers four principal advantages:

\subsubsection{Disparity Disentanglement} Skim Transformers systematically disentangle the process of disparity modeling via a divide-and-conquer strategy. It splits the task into multiple DSA branches, each serving as a specialized prior and constructing its own attention score matrix over a selectively skimmed SAI subset in the query and key matrices to target specific disparity ranges. By ensuring each branch operates separately, any specific disparity, especially those underrepresented in training samples yet critical for high-fidelity reconstructions, is preserved from being overlooked and suppressed. This architecture enables SkimLFSR to become inherently disparity-aware. Further analysis in Sections \ref{sec:DSA_visualisation} and \ref{sec:tSNE_visualisation} investigates how this design correlates with scene depth and camera configurations in LF images. Although these two modalities are not explicitly provided during regression-based training, SkimLFSR appears to develop a discriminative tendency aligned with them.

\subsubsection{Flexibility in Skimmed SAI Sets} Skim Transformers offer flexibility in the construction of skimmed SAI sets, enabling customized attention for any desired disparity range. As illustrated in Fig.\ref{fig:SkimTransformers}, the $DSA_1$ branch leverages the outer SAIs at the corners to model larger disparities, while $DSA_2$ employs inner SAIs for shorter disparity modeling. Both the number and spatial configuration of SAIs can vary between branches and may overlap. Regardless of the configuration, the resulting tokens remain in the spatial domain $(W \times H)$, ensuring that the computed attention matrices maintain a shape of $(WH \times WH)$ before being applied to the value matrix. This flexibility allows the network to meet diverse requirements and scenarios. In Section \ref{sec:Skimmed_SAI_sets}, we experiment with a variety of SAI combinations to showcase this flexibility and analyze the rationale behind SAI selection in maximizing the performance.

\subsubsection{Lower Computational Complexity} Skim Transformers significantly reduce computational redundancy without information loss. This is achieved by applying the skimming process only to the query $\mathcal{Q}_i$ and key $\mathcal{K}_i$ matrices to compute the attention score matrix, while the value matrix $\mathcal{V}_i$ retains the full set of SAIs. As a result, the revised self-attention mechanism still preserves full access to LF information while substantially lowering computational cost. In other words, the principle of ``less'' is realized not at the expense of losing information.

Table \ref{tab:computation} presents a theoretical computation complexity comparison between a Skim Transformer ($\mathcal{O}(SkimT)$) and a conventional spatial Transformer ($\mathcal{O}(T)$) \cite{congLFDET_TMM2024} of four common components, namely matrix projection, attention score matrix, self-attention output, and feed-forward network. Assuming $U=V=5$, $\mathcal{S}_i=4$, $C_D = 2C$, $C_{QK} = C$, and $N_{DSA} = 2$ (as used in the experiments), the Skim Transformer substantially brings computation complexity down to 53\% for matrix projection, 8\% for attention score matrix, and 0\% for feed-forward network, while maintaining the same complexity for self-attention output. A detailed explanation is provided in the supplementary material.

These improvements will be further validated in Section \ref{section:model_efficiency} and Fig. \ref{fig:psnr2efficiency_all}, where the Skim Transformer is shown to achieve a better trade-off between performance and efficiency.

\begin{highlight}
\subsubsection{Cross-angular-resolution Generalization} \label{sec:cross_angular_resolution_generalization} In the LF community, the ability to generalize across different angular resolutions has been relatively underexplored. One key reason is that some existing approaches adopt an explicit encoding of the entire angular space into a fixed-size feature space. For instance, the spatial Transformers used in M2MT-Net \cite{hu_M2MTNet_TMM2024} treat the spatial locations as tokens and apply a linear layer to aggregate angular information from all SAIs, i.e., $\mathbb{R}^{U \times V} \mapsto \mathbb{R}^{C_D}$. Similar designs are also adopted in the convolution-based DistgSSR \cite{wangDistgSSR_TPAMI2023} and HLFSR \cite{duongHLFSR_TCI2023} where their EPI feature extractors employ a convolution kernel of size related to $U$ or $V$, into feature channels $C_D$, and their angular feature extractors further use a convolution kernel of size $(U \times V)$ to aggregate angular information from all the SAIs. These designs inherently couple the model to a specific angular resolution, since $(U \times V)$ is fixed once a model is trained, thereby preventing operation on a different angular resolution.
\end{highlight}

\input{tex/tables/OverallComparison}
\input{tex/figures/OverallComparison}

\begin{highlight}
In contrast, our Skim Transformer is angular-resolution-agnostic, because the disparity embedding process operates not on the complete set of SAIs but on a skimmed subset $\{\mathcal{S}_i\} \subseteq \{U \times V\}$, i.e. $\mathbb{R}^{\mathcal{S}_i} \mapsto \mathbb{R}^{C_D}$ as indicated as $D_i$ in Equation \ref{eq:disparity_embedding}. The model can naturally generalize to a new resolution $U' \times V'$ as long as the skimmed SAI set fits within it, i.e. $\{\mathcal{S}_i\} \subseteq \{U' \times V'\}$. Consequently, this design not only disentangles disparity information itself as described earlier but also disentangles the model from any specific angular resolution, guiding it to learn angular-resolution-agnostic disparity information and achieve a better generalization capability.

This advantage will be validated in Section \ref{sec:larger_angular_resolution} and Table \ref{tab:overall_7x7} where the SkimLFSR model trained on $5 \times 5$ SAIs is tested on $7 \times 7$ SAIs without any retraining and major network modifications.
\end{highlight}

%% file: tex/figures/network.tex
\begin{figure}[t!]
    \centering
    \includegraphics[width=0.48\textwidth]{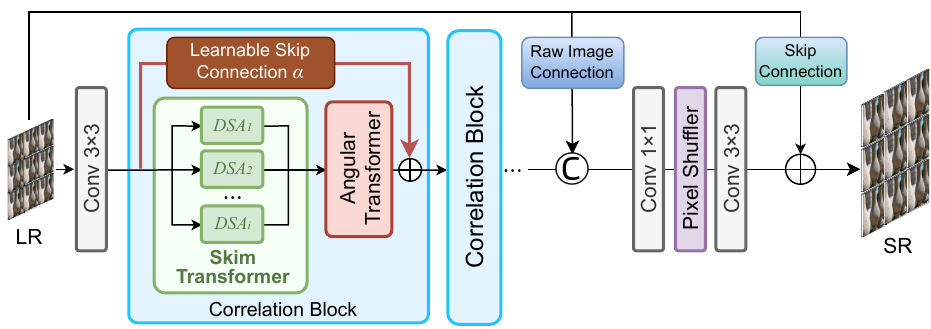}
    \caption{Illustration of the SkimLFSR network.}
    \label{fig:Network}
    \vspace{-10pt}
\end{figure}

%% file: tex/figures/SkimTransformer.tex
\begin{figure*}[ht]
    \centering
    \includegraphics[width=0.80\textwidth]{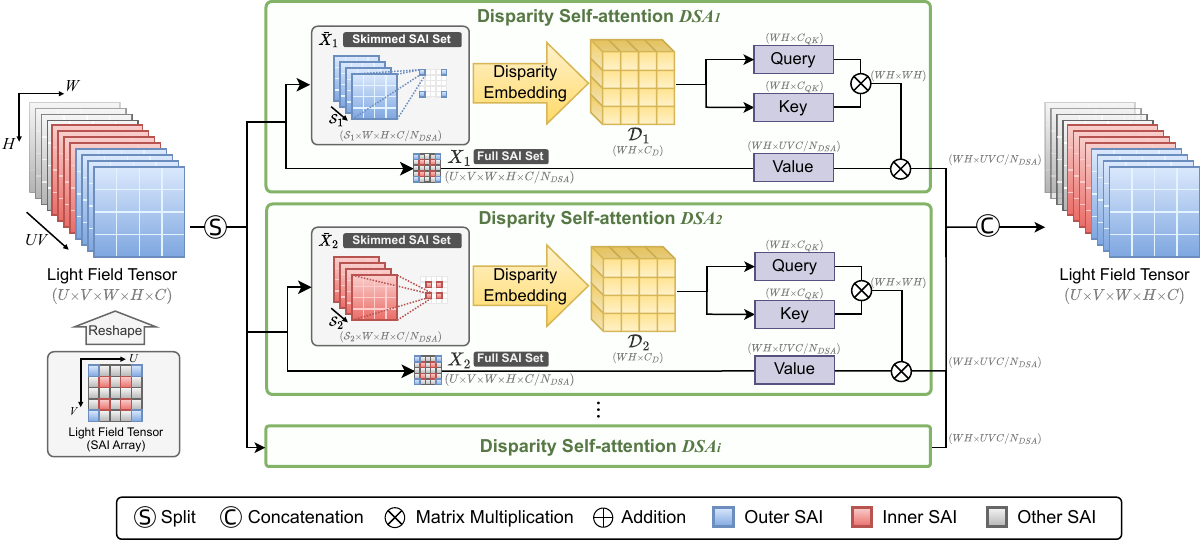}
    \caption{Illustration of Skim Transformer.}
    \label{fig:SkimTransformers}
    \vspace{-5pt}
\end{figure*}

%% file: tex/tables/computation.tex
\begin{table*}[th]
    \caption{Comparison of theoretical computation complexity between a conventional spatial Transformer and a Skim Transformer.}
    \centering
    \label{tab:computation}
    \tabcolsep=0.05cm
    \renewcommand{\arraystretch}{2.0}
    \resizebox{0.99\textwidth}{!}{
        \begin{threeparttable}
        \begin{tabular}{|l|c|c|c|c|}
            \hline
                                                                & $\mathcal{O}(T)$          & $\mathcal{O}(SkimT)$                                                      & $\mathcal{O}(SkimT)/\mathcal{O}(T)$ [Symbolic]                                & $\mathcal{O}(SkimT)/\mathcal{O}(T)$ [Numeric] \\\hline
            (1) Matrix Projection                               & $3UVWHC^2$                & $\sum_{i=1}^{N_{DSA}}\mathcal{S}_iWH(\frac{C}{N_{DSA}}C_D+2C_DC_{QK})$    & $\frac{\sum_{i=1}^{N_{DSA}}\mathcal{S}_i(C_DC/N_{DSA}+2C_DC_{QK})}{3UVC^2}$   & $40:75 \approx 53\%$\\\hline
            (2) Attention Score Matrix                          & $UV(WH)^2C$               & $N_{DSA}(WH)^2C_{QK}$                                                     & $\frac{N_{DSA}C_{QK}}{UVC}$                                                   & $2:25 \approx 8\%$\\\hline
            (3) Self-attention Output                           & $UV(WH)^2C$               & $UV(WH)^2C$                                                               & $1:1$                                                                         & $1:1 = 100\%$ \\\hline
            (4) Feed-forward Network                            & $UVWHC^2$                 & $0$                                                                       & $0:1$                                                                         & $0:1 = 0\%$ \\\hline
        \end{tabular}
        \end{threeparttable}
    }
    \vspace{-5pt}
\end{table*}

%% file: tex/tables/OverallComparison.tex
\begin{table*}[ht!]
\caption{Comparison of PSNR/SSIM. The best and second-best results are highlighted in bold and underlined, respectively.}
\centering
\resizebox{0.90\textwidth}{!}{
\begin{threeparttable}

\begin{tabular}{|l|c|c|c|c|c|c|}
    \hline
    \makecell[l]{Methods\\(\#Train/\#Test samples)} & \makecell{\textit{EPFL}\\(70/10)} & \makecell{\textit{HCInew}\\(20/4)} & \makecell{\textit{HCIold}\\(10/2)} & \makecell{\textit{INRIA}\\(35/5)} & \makecell{\textit{STFgantry}\\(9/2)} & Average \\
    \hline\hline 
    \multicolumn{7}{|l|}{{$\pmb{2 \times} \; \bf{LFSR}$}} \\\hline
    LFSSR-SAV \cite{chengLFSSRSAV_TCI2022}      & 34.62/0.9772                          & 37.42/0.9776                          & 44.22/0.9942                              & 36.36/0.9849                            & 38.69/0.9914                            & 38.26/0.9851                                  \\
    DKNet \cite{huDKNet_TIM2022}                & 34.01/0.9759                          & 37.36/0.9780                          & 44.19/0.9942                              & 35.80/0.9843                            & 39.59/0.9910                            & 38.19/0.9847                                  \\
    DPT \cite{wangDPT_AAAI2022}                 & 34.49/0.9758                          & 37.36/0.9771                          & 44.30/0.9943                              & 36.41/0.9843                            & 39.43/0.9926                            & 38.40/0.9848                                  \\
    DistgSSR \cite{wangDistgSSR_TPAMI2023}	    & 34.81/0.9787                          & 37.96/0.9796                          & 44.94/0.9949                              & 36.58/0.9859                            &	40.40/0.9942                            & 38.94/0.9867                                  \\
    LFT \cite{liangLFT_SPL2022}	                & 34.78/0.9776                          & 37.77/0.9788                          & 44.63/0.9947                              & 36.54/0.9853                            &	40.41/0.9941                            & 38.82/0.9861                                  \\
    HLFSR \cite{duongHLFSR_TCI2023}	            & 35.31/0.9800                          & 38.32/0.9807                          & 44.98/0.9950                              & 37.06/0.9867                            &	40.85/0.9947                            & 39.30/0.9874                                  \\
    EPIT \cite{liangEPIT_ICCV2023}	            & 34.85/0.9775                          & 38.23/\underline{0.9810}              & 45.08/0.9949                              & 36.68/0.9852                            &	\underline{42.17}/\underline{0.9957}    & 39.40/0.9869                                  \\
    LF-DET \cite{congLFDET_TMM2024}	            & 35.20/0.9794                          & 38.22/0.9803                          & 44.92/0.9949                              & 36.88/0.9862                            &	41.56/0.9953                            & 39.36/0.9872                                  \\
    LFMamba \cite{xia_LFMamba_arxiv2024}	    & \underline{35.76}/\underline{0.9824}  & 38.37/0.9801                          & 44.99/0.9950                              & 37.06/\underline{0.9876}                &	40.95/0.9948                            & 39.42/\underline{0.9881}                      \\
    M2MT-Net \cite{hu_M2MTNet_TMM2024}          & 35.64/0.9815                          & \underline{38.43}/\underline{0.9810}  & \underline{45.38}/\underline{0.9953}      & \underline{37.22}/0.9870                & 40.99/0.9949                            & \underline{39.53}/0.9879                      \\
    \hline
    \multirow{2}{*}{SkimLFSR (Ours)}            & \textbf{36.18}/\textbf{0.9836}        & \textbf{38.89}/\textbf{0.9825}        & \textbf{45.62}/\textbf{0.9955}            & \textbf{37.60}/\textbf{0.9881}          &	\textbf{42.52}/\textbf{0.9962}          & \textbf{40.16}/\textbf{0.9892}                \\
                                                & \green{+0.42}/\green{+0.0012}         & \green{+0.46}/\green{+0.0015}         & \green{+0.24}/\green{+0.0002}             & \green{+0.38}/\green{+0.0005}           & \green{+0.35}/\green{+0.0005}           & \green{+0.63}/\green{+0.0009}                 \\
    \hline\hline
    \multicolumn{7}{|l|}{{$\pmb{4 \times} \; \bf{LFSR}$}} \\\hline
    LFSSR-SAV \cite{chengLFSSRSAV_TCI2022}      & 29.37/0.9223                          & 31.45/0.9217                          & 37.50/0.9721                              & 31.27/0.9531                            & 31.36/0.9505                            & 32.19/0.9439                                  \\
    DKNet \cite{huDKNet_TIM2022}                & 28.85/0.9174                          & 31.17/0.9185                          & 37.31/0.9720                              & 30.80/0.9501                            & 30.85/0.9460                            & 31.80/0.9408                                  \\
    DPT \cite{wangDPT_AAAI2022}                 & 28.94/0.9170                          & 31.20/0.9188                          & 37.41/0.9721                              & 30.96/0.9503                            & 31.15/0.9488                            & 31.93/0.9414                                  \\
    DisgSSR \cite{wangDistgSSR_TPAMI2023}       & 28.99/0.9195                          & 31.38/0.9217                          & 37.56/0.9732                              & 30.99/0.9519                            & 31.65/0.9534                            & 32.12/0.9439                                  \\
    LFT \cite{liangLFT_SPL2022}                 & 29.33/0.9196                          & 31.36/0.9205                          & 37.59/0.9731                              & 31.30/0.9515                            & 31.62/0.9530                            & 32.24/0.9436                                  \\
    HLFSR \cite{duongHLFSR_TCI2023}             & 29.20/0.9222                          & 31.57/0.9238                          & 37.78/0.9742                              & 31.24/0.9534                            & 31.64/0.9537                            & 32.28/0.9455                                  \\
    EPIT \cite{liangEPIT_ICCV2023}              & 29.31/0.9196                          & 31.51/0.9231                          & 37.68/0.9737                              & 31.35/0.9526                            & 32.18/0.9570                            & 32.41/0.9452                                  \\
    LF-DET \cite{congLFDET_TMM2024}	            & 29.42/0.9220                          & 31.51/0.9227                          & 37.76/0.9739                              & 31.34/0.9528                            &	32.02/0.9561                            & 32.41/0.9455                                  \\
    LFMamba \cite{xia_LFMamba_arxiv2024}	    & 29.83/0.9257                          & 31.70/0.9249                          & 37.91/0.9748                              & \underline{31.81}/0.9551                &	31.85/0.9554                            & 32.62/0.9472                                  \\
    M2MT-Net \cite{hu_M2MTNet_TMM2024}	        & \underline{29.85}/\underline{0.9284}  & \underline{31.76}/\underline{0.9264}  & \underline{37.98}/\underline{0.9749}      & 31.77/\underline{0.9563}                &	\underline{32.20}/\underline{0.9584}    & \underline{32.71}/\underline{0.9489}          \\
    \hline
    \multirow{2}{*}{SkimLFSR (Ours)}            & \textbf{30.06}/\textbf{0.9308}        & \textbf{32.03}/\textbf{0.9289}        & \textbf{38.26}/\textbf{0.9762}            & \textbf{31.92}/\textbf{0.9572}          &	\textbf{33.00}/\textbf{0.9630}          & \textbf{33.06}/\textbf{0.9512}                \\
                                                & \green{+0.21}/\green{+0.0024}         & \green{+0.27}/\green{+0.0025}         & \green{+0.28}/\green{+0.0013}             & \green{+0.11}/\green{+0.0009}           & \green{+0.80}/\green{+0.0046}           & \green{+0.35}/\green{+0.0023}                 \\\hline
\end{tabular}
\end{threeparttable}
}
\label{tab:overall}
\vspace{-3pt}
\end{table*}

%% file: tex/figures/OverallComparison.tex
\begin{figure}[ht!]
    \centering
    \tabcolsep=0.01cm
    \begin{tabular}{cc}
        (a) $2 \times$ LFSR &
        (b) $4 \times$ LFSR \\
        \includegraphics[width=0.24\textwidth]{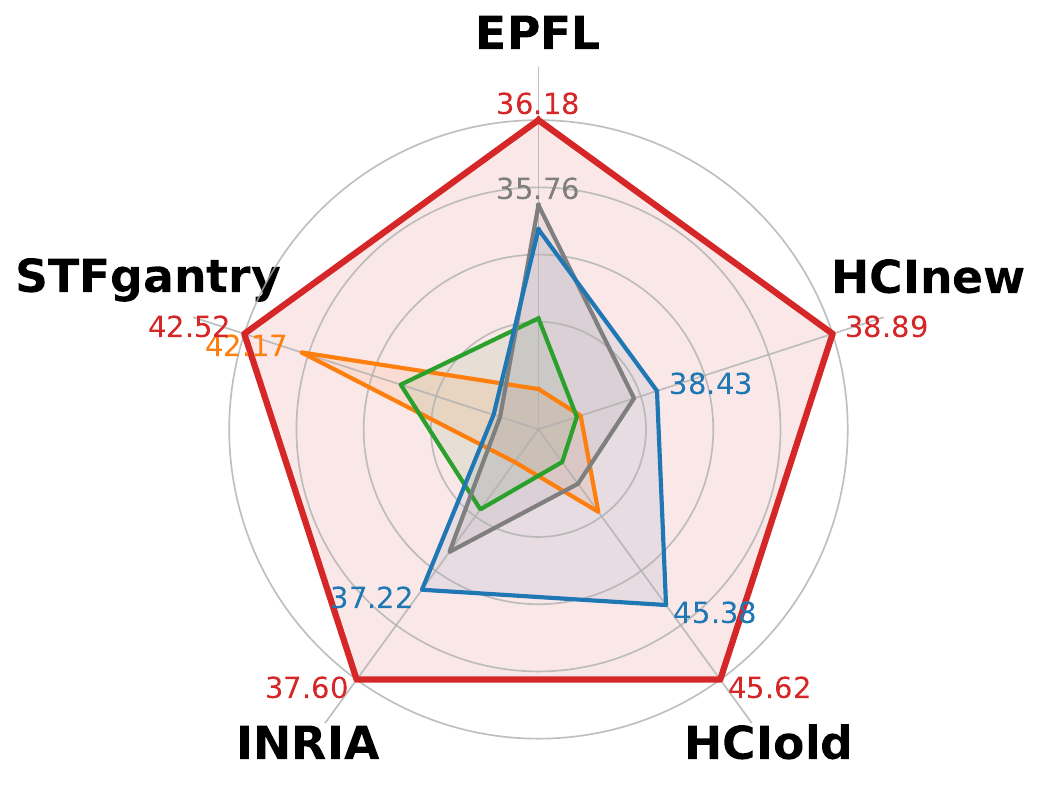} &
        \includegraphics[width=0.24\textwidth]{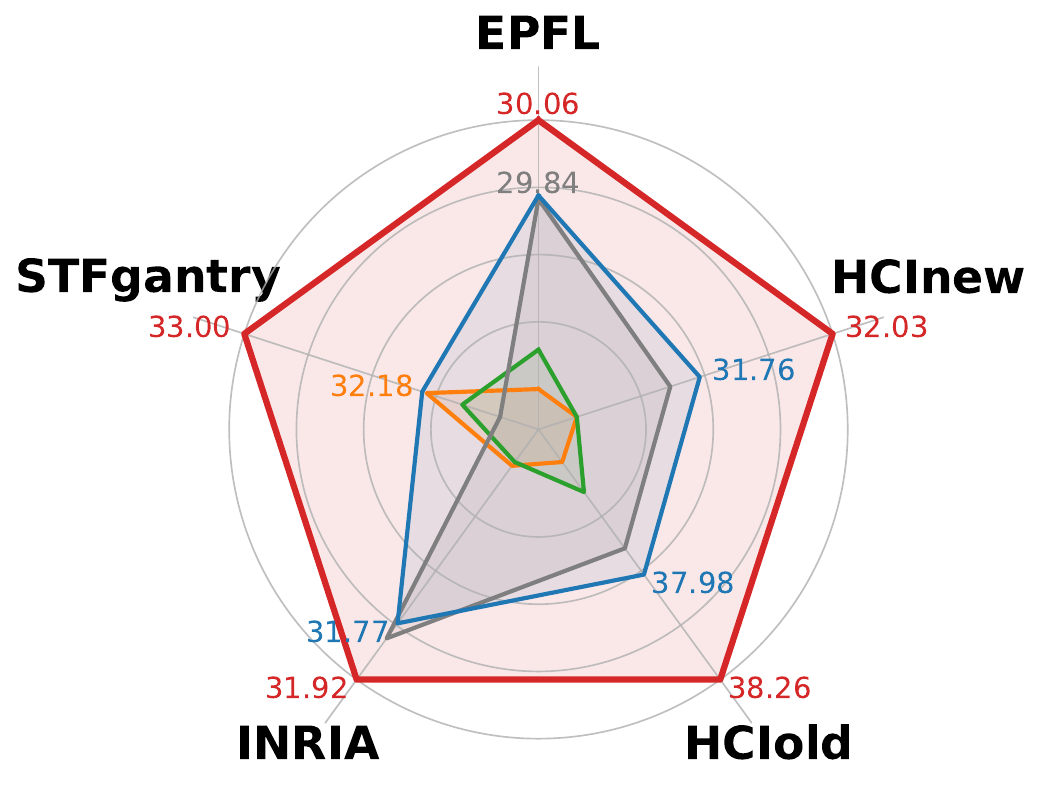} \\
        \multicolumn{2}{c}{\includegraphics[width=0.32\textwidth]{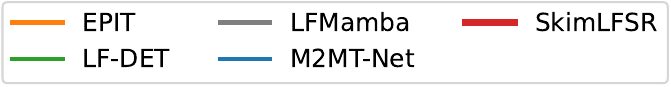}}
    \end{tabular}
    \caption{Comparison of PSNR on the five testing datasets.}
    \label{fig:OverallComparison}
    \vspace{-5pt}
\end{figure}

%% file: tex/4.Experiments.tex
\section{Experiments}
\subsection{Experimental Settings}
We conduct experiments using the widely adopted BasicLFSR framework \cite{BasicLFSR} and following its protocol. It consists of five datasets: \textit{EPFL} \cite{rerabekEPFL2016}, \textit{HCInew} \cite{honauerHCInew_ACCV2016}, \textit{HCIold} \cite{wannerHCIold_VMV2013}, \textit{INRIA} \cite{lependuINRIA_TIP2018}, and \textit{STFgantry} \cite{vaishSTFgantry_2008}. These datasets are split into 70/20/10/35/9 training samples and 10/4/2/5/2 testing samples, respectively. SkimLFSR is trained for 100 epochs with a learning rate of $2 \times 10^{-4}$, followed by an additional 10 epochs for fine-tuning with a learning rate of $2 \times 10^{-5}$.

Regarding SkimLFSR's hyperparameters, we empirically set the feature channels $C = 48$ across all Transformers and convolutions, and $C_D = 96$ and $C_{QK} = 48$ for Skim Transformers. Within each Skim Transformer, as shown in Fig. \ref{fig:SkimTransformers}, we employ two DSA branches ($N_{DSA} = 2$), one of which leverages the four outer SAIs (blue) representing the largest disparity range, and the other one uses the four inner SAIs (red) closer to the center for the smaller disparity range. Due to the improved efficiency of Skim Transformers, we can construct a deeper network with 20 correlation blocks ($N_{CB} = 20$) for both the $2 \times$ and $4 \times$ LFSR scale.

The experiments are conducted on a workstation equipped with an Intel i7-11700 4.80GHz 8-core CPU, 32 GB RAM, and an NVIDIA RTX 3090 GPU. The implementation code and trained models will be publicly available at \href{https://huzexi.github.io/}{https://huzexi.github.io/}.

\subsection{Quantitative Comparison}
A quantitative comparison is conducted to compare the overall performance of SkimLFSR with state-of-the-art methods at the $2 \times$ and $4 \times$ scales. As per the protocol \cite{BasicLFSR}, only the central $5 \times 5$ SAIs are used. PSNR and SSIM are calculated on the \textit{Y} channel of the \textit{YCbCr} color space as the quantitative metrics of model performance.

The comparative evaluation encompasses both convolution-based, Transformer-based, \hl{and Mamba-based methods}. The convolution-based methods include LFSSR-SAV \cite{chengLFSSRSAV_TCI2022}, DKNet \cite{huDKNet_TIM2022}, DistgSSR \cite{wangDistgSSR_TPAMI2023}, and HLFSR \cite{duongHLFSR_TCI2023}. The Transformer-based methods comprise DPT \cite{wangDPT_AAAI2022}, LFT \cite{liangLFT_SPL2022}, EPIT \cite{liangEPIT_ICCV2023}, LF-DET \cite{congLFDET_TMM2024}, and M2MT-Net \cite{hu_M2MTNet_TMM2024}. \hl{The Mamba-based methods include LFMamba \cite{xia_LFMamba_arxiv2024}.}

\input{tex/figures/qualitative.tex}

The quantitative results presented in Table \ref{tab:overall} demonstrate that SkimLFSR achieves superior performance across both $2\times$ and $4\times$ scales across all datasets. Specifically, SkimLFSR attains an average performance advantage of 0.63 dB and 0.35 dB PSNR over the second-best M2MT-Net at $2\times$ and $4\times$ scales, respectively. When examining individual datasets, the performance gains range from a minimum of 0.24 dB PSNR on the \textit{HCIold} dataset at the $2\times$ scale and 0.11 dB PSNR on \textit{INRIA} at the $4\times$ scale. A particularly strong performance is observed on the \textit{EPFL} dataset at the $2\times$ scale, with SkimLFSR achieving a gain of 0.42 dB, and an even greater improvement of 0.80 dB on the \textit{STFgantry} dataset at the $4 \times$ scale. This substantial improvement aligns with our hypothesis that the disparity disentanglement capability of Skim Transformers enables SkimLFSR to effectively handle diverse disparity ranges and deliver a balanced performance, and this advantage is particularly pronounced on the severely underrepresented \textit{STFgantry} dataset as it features larger disparity ranges but contains only nine training samples, representing merely 6.25\% (9 out of 144) of the training samples, making it the most data-scarce subset in our evaluation.

\hl{To provide additional insight into SkimLFSR's comparative advantages, we present a radar plot in Fig. \ref{fig:OverallComparison}. At the $2\times$ scale, although EPIT exhibits strong performance on the large-disparity \textit{STFgantry} dataset, this appears to result from overfitting to large disparity ranges, as evidenced by its substantial performance degradation on datasets featuring smaller disparity ranges, such as \textit{EPFL} and \textit{INRIA}. Mixed performance is also observed for M2MT-Net and LFMamba, which achieve strong results on \textit{EPFL} and \textit{INRIA} but underperform on \textit{STFgantry}. In comparison, our SkimLFSR achieves a consistent performance advantage across all datasets, showing no discernible shortcomings on either large- or small-disparity datasets.}

At the $4\times$ scale, recent methods yield moderate improvements over earlier approaches across most datasets. Notably, \textit{STFgantry} represents a clear exception to this trend, where the best compared method, M2MT-Net, achieves only a negligible improvement of 0.02 dB PSNR over the earlier method EPIT. This stark contrast highlights the fundamental challenge faced by existing LFSR networks in maintaining consistently strong performance across varying disparity ranges. In contrast, SkimLFSR achieves a sweeping win across all datasets, with a particularly significant gain of 0.80 dB PSNR on the challenging \textit{STFgantry} dataset. This comprehensive and consistent superiority underscores SkimLFSR's exceptional robustness and effectiveness in handling diverse disparity conditions.

\subsection{Qualitative Comparison}
We showcase SkimLFSR's superior performance through qualitative evaluations at the $4 \times$ scale in Fig. \ref{fig:Qual}.

The first example, shown in (a) \textit{Perforated\_Metal\_3}, is a challenging scene that features complex occlusion patterns with dense perforations and bright light spots. SkimLFSR's reconstruction exhibits remarkable detail preservation, as evidenced in the blue region where light spot details emerge clearly, and in the red region where both light spots and hole edges are sharply delineated. In contrast, competing methods suffer from light leakage artifacts around bright spots and produce indistinct hole boundaries. The corresponding EPIs further confirm that SkimLFSR maintains superior detail fidelity and angular consistency compared to other methods.

Similar improvements are observed in (b) \textit{Building\_Decoded} from the INRIA dataset. The error maps within the highlighted regions reveal that this performance enhancement stems from more accurate edge reconstruction and reduced artifacts along building window boundaries. The zoom-in views and EPIs corroborate these findings, showing notably sharper window edges in SkimLFSR's output.

(c) \textit{Lego Knights} is a synthetic sample from the \textit{STFgantry} dataset, where SkimLFSR achieves the most substantial improvement of 1.12 dB PSNR over the second-best EPIT. The reconstruction quality differences are particularly evident in the slim lines in the background. The blue box highlights one of these regions. This exemplifies SkimLFSR's capability to accurately handle fine details with small disparities in the background. The foreground reconstruction also benefits from SkimLFSR's ability, as depicted in the red box, where the stud edges are reconstructed with substantially lower error magnitudes in SkimLFSR's error map.

These examples collectively validate SkimLFSR's exceptional ability to reconstruct fine details across diverse disparity ranges and structural characteristics, yielding consistent performance advantages in both foreground and background regions.

\input{tex/figures/psnr2efficiency}

\subsection{Model Efficiency} \label{section:model_efficiency}

We assess the efficiency of SkimLFSR by benchmarking it against state-of-the-art methods at the $4 \times$ scale. To assess performance across a spectrum of model complexities, we vary the number of correlation blocks $N_{CB}$ from 4 to 20 in increments of two. The compared methods include the Transformer-based LFT, EPIT, LF-DET, and M2MT-Net, as well as the convolution-based DistgSSR,  LFSSR-SAV, and HLFSR. The results are plotted in Fig. \ref{fig:psnr2efficiency_all}. In the plots, the average PSNR is used as the performance metric on the vertical axis. Three efficiency metrics are involved as the horizontal axis: (a) parameter number, (b) FLOPs, and (c) inference time. Ideally, a method positioned closer to the top-left corner of each plot is considered more efficient, as it achieves higher performance with fewer computational resources. The detailed numerical results are provided in the supplementary material.

Our SkimLFSR network consistently excels across all three efficiency metrics. Notably, the most lightweight variant of SkimLFSR with $N_{CB} = 4$ has already achieved a PSNR of 32.47 dB, outperforming all compared methods except M2MT-Net, and surpassing LF-DET by 0.06 dB. Remarkably, it does so with only 37\% of the parameters, 35\% of the FLOPs, and 28\% of the inference time compared to LF-DET. This result aligns with our analysis in Section \ref{sec:Discussion} and Table \ref{tab:computation}, validating that through the skimming strategy, Skim Transformers reduces computational redundancy while maintaining information integrity and emphasizing more pertinent information, embodying the ``less is more'' philosophy.

Compared to the best existing method, M2MT-Net, the smallest SkimLFSR variant that surpasses it is with $N_{CB} = 8$, achieving 32.80 dB, which is 0.09 dB higher. While the FLOPs and inference time are comparable (97\% and 108\%, respectively), this SkimLFSR model requires only 28\% of the parameters, leading to a significantly lighter architecture. The reduced model size not only lowers the risk of overfitting but also enables easier scalability to deeper networks. \hl{Meanwhile, SkimLFSR also surpasses the Mamba-based LFMamba, and it requires only 49\% of the parameters and 45\% of the FLOPs, and is 19\% faster in inference time.}

Indeed, as $N_{CB}$ increases, SkimLFSR consistently improves in performance. The variant with $N_{CB}=20$ reaches 33.06 dB, outperforming M2MT-Net and LF-DET by 0.35 dB and 0.65 dB, respectively. Remarkably, this model still uses only 67\% of the parameters of M2MT-Net, and is only 4\% slower than LF-DET \hl{and 19\% faster than LFMamba} in inference time, while delivering significantly better results. \hl{After this point, the model performance starts to drop when adding more correlation blocks ($N_{CB}=22$ and $24$), indicating that the model has reached capacity saturation.}

In summary, these results demonstrate that SkimLFSR achieves a compelling balance between performance and efficiency. It is capable of delivering lightweight variants, such as $N_{CB} = 4$ and $N_{CB} = 8$, that offer strong performance under limited computational budgets. At the same time, it provides scalable configurations such as $N_{CB} = 20$, which significantly outperform existing methods while maintaining a manageable model size and runtime. This adaptability makes SkimLFSR a practical and powerful solution for a wide range of real-world and high-demand LFSR applications.

\input{tex/figures/dsa_features.tex}
\input{tex/figures/DDR.tex}

\subsection{In-depth Analysis}
To better understand the underlying mechanism driving SkimLFSR's superior performance, we further conduct a series of in-depth analyses on the challenging $4\times$ scale.
\subsubsection{DSA Feature Visualization} \label{sec:DSA_visualisation}
We first analyze the internal behavior of the Skim Transformer by visualizing the output features of its DSA modules. Fig. \ref{fig:DSA_features} displays the feature maps produced by $DSA_1$ and $DSA_2$ (first two columns), along with their differential activations ($DSA_1 - DSA_2$ and $DSA_2 - DSA_1$ in the last two columns). These differential maps reveal the exclusive focus areas of each DSA branch. The visualized samples are the same as those in Fig. \ref{fig:Qual}.

A general trend is observed that, while both DSA branches exhibit overlapping activation regions, examination of the differential maps clearly demonstrates that the two branches possess distinct tendencies for discerning different disparity ranges and attending to complementary aspects of LF images.

Specifically, in (a) \textit{Perforated\_Metal\_3}, $DSA_1 - DSA_2$ demonstrates strong activation on the perforated metal surface in the foreground, whereas $DSA_2 - DSA_1$ predominantly highlights the background regions behind the metal board. This disparity separation is particularly noteworthy since the foreground and background regions exhibit markedly different disparity ranges, suggesting that SkimLFSR has developed an inherent capability to distinguish the perforated metal structure from the background.

The result in (b) \textit{Building\_Decoded} reveals a different specialization pattern. Here, $DSA_1 - DSA_2$ appears to attend preferentially to background regions, including building walls, sky regions, and grass fields. Conversely, $DSA_2 - DSA_1$ exhibits a pronounced tendency to focus on structural boundaries, particularly edges of the building and its windows.

Interestingly, an opposite pattern is observed in (c) \textit{Lego Knights}. In this case, $DSA_1 - DSA_2$ primarily attends to structural edges, such as the knights and studs, while $DSA_2 - DSA_1$ focuses on background regions. This reversal can be attributed to the fundamental differences between samples (b) and (c): the former originates from a Lytro Illum camera capturing a far scene, while the latter is synthetically generated with a completely different camera configuration and presents near objects in a simulated environment. As a result, these two DSA branches demonstrate adaptive behavior, attending to content with different characteristics as well as the specific camera configuration.

Notably, despite being trained solely on the regression-based LFSR task, considering separation of the perforated surface in (a) \textit{Perforated\_Metal\_3} and locating the structures of building in (b) \textit{Building\_Decoded} and lego in (c) \textit{Lego Knights}, SkimLFSR demonstrates emergent behavior resembling classification, suggesting implicit learning of semantics akin to classification tasks, like depth estimation \cite{wangOcclusionawareDepthEstimation2015} and shape extraction \cite{heberUshapeICCV2017}, highlighting that SkimLFSR has developed adaptive disparity-aware capabilities that facilitate comprehensive scene understanding during reconstruction.

\subsubsection{t-SNE Visualization of DSA Features} \label{sec:tSNE_visualisation}
We further analyze SkimLFSR’s feature space using deep degradation representation (DDR), a technique originally proposed for interpreting single image super-resolution (SISR) models. In prior work \cite{liuDDR_arxiv2022, kong_reflash_CVPR2022}, it is discovered that by applying PCA \cite{hotelling_PCA_JEP1933} and t-SNE \cite{liuEvaluatingGeneralization_TPAMI2023} to project the deep features into a two-dimensional space, well-trained SISR methods exhibit a clustering trend based on degradation types, suggesting that these methods have implicitly learned degradation as semantics, even though such information is not explicitly provided during training.

Inspired by this observation, we hypothesize that an effective LFSR method should similarly exhibit patterns related to disparities. To validate it, we perform the same technique on the three LFSR methods: EPIT \cite{liangEPIT_ICCV2023}, LF-DET \cite{congLFDET_TMM2024}, and M2MT-Net \cite{hu_M2MTNet_TMM2024}, on all the 167 samples, as shown in Fig. \ref{fig:DDR}. These samples are categorized into three camera configurations: the Lytro Illum camera group (blue), \textit{EPFL} \cite{rerabekEPFL2016}, \textit{INRIA} \cite{lependuINRIA_TIP2018}; the synthetic group (green), \textit{HCInew} \cite{honauerHCInew_ACCV2016}, \textit{HCIold} \cite{wannerHCIold_VMV2013}; and the gantry-based camera group (red), \textit{STFgantry} \cite{vaishSTFgantry_2008}. Each group is characterized by distinct disparity distributions due to different imaging setups: the Lytro Illum and gantry-based cameras employ different microlens configurations, while synthetic datasets simulate yet another unique angular distribution. The Calinski-Harabasz Index (CHI) \cite{calinski_CHI_1974} is computed for each method to quantify the clustering trend, with higher scores indicating better separability of the clusters.

Despite the clear disparity distinctions across camera configuration groups, the compared methods exhibit no evident clustering trend, suggesting a lack of disparity-aware representation. LF-DET demonstrates the weakest separation (CHI: 6.48), with all three groups dispersed across the feature space. EPIT (CHI: 17.25) and M2MT-Net (CHI: 11.86) display a slight clustering trend, but substantial overlaps persist, with numerous outliers from each group intermingled with others. These results point to the potential disparity entanglement issue in the learned feature representations of these methods.

\input{tex/tables/dsa_variants.tex}
\input{tex/figures/dsa_12_vs_dsa_full.tex}

In contrast, SkimLFSR exhibits a clear and well-defined clustering pattern, with distinct boundaries separating the three groups and only minor overlaps. Its CHI score of 41.27 is substantially higher than those of the other methods, indicating that SkimLFSR has effectively learned to organize features based on disparity information.

This outcome strongly echoes our earlier observations in Fig. \ref{fig:DSA_features}, where SkimLFSR demonstrates classification-like behavior despite being trained solely on a regression-based LFSR task. Whereas the previous analysis highlights SkimLFSR’s ability to distinguish scene depths within a single LF image, the current t-SNE study reveals its capacity to differentiate camera configurations across LF images. Since scene depth and camera configuration are the two primary factors influencing disparity variations, these studies collectively prove that SkimLFSR has developed a latent discriminative capability toward disparity information, i.e., disparity-aware, which finally contributes to its superior performance in LFSR.

\subsubsection{Ablation study on Skimmed SAI Sets} \label{sec:Skimmed_SAI_sets}

The configuration of the skimmed SAI sets in the DSA branches serves as critical prior knowledge for targeted disparity modeling in the Skim Transformer. To evaluate its impact, we conduct an ablation study, with results presented in Table \ref{tab:DSA_variants}. We adopt the $N_{CB}=6$ model as the baseline, representing the full version of Skim Transformers in Variant (a).

Variants (b) and (c) isolate the two individual skimmed SAI sets (blue and red) used in the baseline. Each subset underperforms the combined variant by 0.11 dB and 0.13 dB, respectively, demonstrating the advantage of combining different SAI sets to capture a broader disparity range.

Variants (d), (e), and (f) explore variants of Variant (a), (b), and (c) using only half the number of SAIs. These configurations, which sample along a single diagonal line with a biased disparity coverage, lead to further degradation in performance: a drop of 0.18 dB for the dual-branch variant (Variant (a) vs. (d)), and 0.28 dB for the single-branch variants (Variant (a) vs. (e) and (f)). These results highlight the importance of both the quantity and spatial distribution of SAIs, as they directly influence the disparity coverage and, consequently, the model’s reconstruction quality.

\input{tex/tables/OverallComparison_7x7.tex}

Variant (g) experiments with a minimal variant using only a single central SAI (yellow). In this setting, the angular scope of the attention score matrix is severely restricted, resulting in a substantial performance drop of 0.53 dB compared to the baseline. This highlights the necessity of the multi-branch disparity modeling.

While the previous cases reduce either the number of DSA branches or the number of SAIs, each causing performance degradation, we find that merely increasing the number of SAIs does not necessarily lead to better results. Variant (h) augments the baseline with a third DSA branch comprising central SAIs from each edge (green). Although this triple-branch variant introduces additional parameters and computational cost, it does not yield any noticeable performance improvement. This outcome suggests that the baseline configuration already provides sufficient disparity range coverage.

In a more extreme case, Variant (i) consolidates all SAIs into a single DSA branch. This effectively reintroduces disparity entanglement back to the model, as the single branch must process all the SAI at once. \hl{Despite a slight performance drop of 0.08 dB and a similar FLOPs and inference time, this all-in-one configuration results in a 61\% increase in model parameters. This increase in parameters not only introduces additional computational burden but also raises the risk of overfitting and compromises the model’s scalability. To further investigate this effect, we compare the baseline Variant (a) with the all-in-one Variant (i) by varying the number of correlation blocks $N_{CB}$ from 4 to 20 in increments of two. The results, shown in Fig. \ref{fig:dsa_12_vs_dsa_full}, reveal that Variant (i) not only underperforms the baseline across all $N_{CB}$ but also saturates early at $N_{CB} = 16$ with only 32.76 dB PSNR, whereas the baseline continues to improve and reaches 33.06 dB at $N_{CB} = 20$.} 

Collectively, these findings confirm that the information structure introduced by the DSA branches is far more important than merely increasing the amount of information processed. A more compact model with fewer parameters not only avoids unnecessary computational and overfitting burdens but also achieves better scalability, embodying the ``less is more'' philosophy.

\begin{highlight}
\subsection{Larger Angular Resolution}
\subsubsection{Qualitative Comparison}
The standard protocol of BasicLFSR \cite{BasicLFSR} only engages the central $5 \times 5$ SAIs. To explore the scalability of SkimLFSR to larger angular resolutions, we conduct an experiment on the $7 \times 7$ SAIs to compare SkimLFSR with M2MT-Net \cite{hu_M2MTNet_TMM2024} and LF-DET \cite{congLFDET_TMM2024}. To handle the larger angular resolution, three DSA branches are employed in the Skim Transformers to handle the three corner SAI sets. The results are presented in the first group of Table \ref{tab:overall_7x7}.

SkimLFSR maintains its superiority over the other methods in this task as in the $5 \times 5$ SAIs, achieving an average PSNR gain of 0.18 dB over the second-best method M2MT-Net. Particularly, on the \textit{EPFL} and \textit{STFgantry} datasets, SkimLFSR achieves a significant PSNR gain of 0.26 dB and 0.28 dB, respectively. This demonstrates that SkimLFSR is capable of handling different disparity ranges in a larger angular resolution.

\subsubsection{Skimmed SAI Sets}
As the angular resolution increases, the number of possible skimmed SAI combinations grows accordingly. As in Section \ref{sec:Skimmed_SAI_sets}, we conduct an ablation study to evaluate the impact of different skimmed SAI sets. Using the three-branch SkimLFSR as the baseline, we construct three two-branch variants, (a), (b), and (c), each formed by removing one branch and thus retaining one of the three possible pairs of corner SAI sets. The results are presented in the second group of Table \ref{tab:overall_7x7}.

Overall, a consistent trend of average performance degradation is observed across the two-branch variants, ranging from 0.08 dB to 0.11 dB, relative to the three-branch baseline. While the overall degradation is limited, larger drops occur on specific datasets. Variants (a) and (c) fall behind by 0.15 dB on \textit{STFgantry}, Variant (b) lags by 0.17 dB on \textit{INRIA}, and Variant (c) shows a 0.18 dB decrease on \textit{EPFL}.

These observations suggest that SkimLFSR benefits from employing more DSA branches when operating at higher angular resolutions, as additional branches provide more coverage of disparity ranges and consequently yield more robust reconstruction performance.

\subsubsection{Cross-angular-resolution Generalization} \label{sec:larger_angular_resolution}
As discussed in Section \ref{sec:cross_angular_resolution_generalization}, SkimLFSR is inherently angular-resolution-agnostic. In fact, many existing LFSR methods are also angular-resolution-agnostic.

We validate this property by directly inferring models trained on the $5 \times 5$ LF data to the $7 \times 7$ LF data, without any retraining or major network modification. LFT \cite{liangLFT_SPL2022}, EPIT \cite{liangEPIT_ICCV2023}, LF-DET \cite{congLFDET_TMM2024}, and LFMamba \cite{xia_LFMamba_arxiv2024} are included in this study. DistgSSR \cite{wangDistgSSR_TPAMI2023}, HLFSR \cite{duongHLFSR_TCI2023} and M2MT-Net \cite{hu_M2MTNet_TMM2024} are not applicable for the reasons mentioned in Section \ref{sec:cross_angular_resolution_generalization}. At the $7 \times 7$ angular resolution, the model has access to a larger angular space. Thus, the two-branch SkimLFSR variant has the freedom to choose among three corner SAI sets to attend to. We therefore construct three variants, (d), (e), and (f), mirroring the same corner SAI sets as Variants (a), (b), and (c) in the previous study. Results are shown in the third group of Table \ref{tab:overall_7x7}.

It is interesting to observe that EPIT, which underperforms many existing methods on the $5 \times 5$ SAIs task in Table \ref{tab:overall}, surpasses them on the $7 \times 7$ SAIs task. In particular, EPIT outperforms the Mamba-based method LFMamba by an average margin of 0.32 dB, whereas it lags behind by 0.21 dB on the $5 \times 5$ setting.

For SkimLFSR, an even more surprising result is observed. Despite never being trained on $7 \times 7$ LF data, the $5 \times 5$ SkimLFSR variant achieves performance that is only up to 0.06 dB lower than the two-branch SkimLFSR variants in the second group that are trained on the $7 \times 7$ LF data, and the improvement of training on $7 \times 7$ mostly comes from the \textit{STFgantry} dataset, bringing the PSNR score from 32.66-32.71 dB to 32.81 to 32.85 dB. It is also noteworthy that the generalized $5 \times 5$ SkimLFSR model still outperforms both LF-DET and M2MT-Net in the first group, even though those methods in the first group are trained on $7 \times 7$ LF data as well.

Motivated by these counter-intuitive observations, we hypothesize that a method’s angular generalizability is not necessarily correlated with its performance at a specific angular resolution. We attribute SkimLFSR’s strong performance in this study to the ability of its DSA branches to capture essential disparity information that is angular-resolution-agnostic, rather than cues that are tightly coupled to a particular angular configuration.

\end{highlight}

%% file: tex/figures/qualitative.tex
\begin{figure*}[ht!]
    \centering
    \tabcolsep=0.05cm
    \renewcommand{\arraystretch}{0.8}
    \resizebox{0.90\textwidth}{!}{
    \begin{tabular}{cccccc}
        % Title
        &
        Ground-truth &
        EPIT &
        LF-DET &
        M2MT-Net &
        SkimLFSR (Ours) \\

        % Perforated_Metal_3
        % main image
        \multirow{4}{*}{\raisebox{10pt}[\height][0pt]{\includegraphics[width=0.180\textwidth, height=0.120\textwidth]{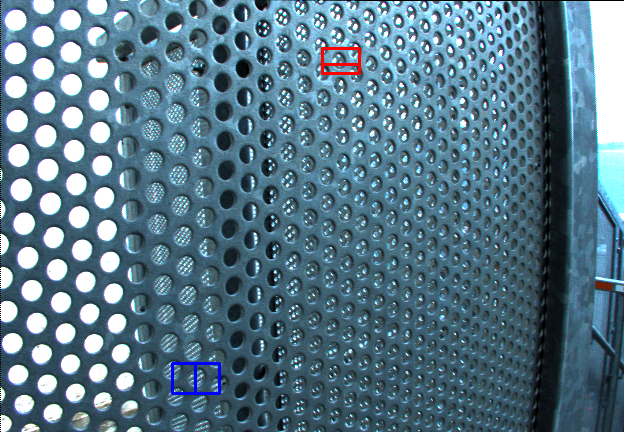}}} &
        % Two images
        \begin{minipage}{0.180\textwidth}
            \includegraphics[width=0.462\textwidth, height=0.28\textwidth,cfbox=blue 1pt 0pt]{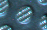}
            \includegraphics[width=0.462\textwidth, height=0.28\textwidth,cfbox=red 1pt 0pt]{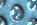}
        \end{minipage} &
        \begin{minipage}{0.180\textwidth}
            \includegraphics[width=0.462\textwidth, height=0.28\textwidth,cfbox=blue 1pt 0pt]{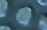}
            \includegraphics[width=0.462\textwidth, height=0.28\textwidth,cfbox=red 1pt 0pt]{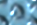}
        \end{minipage} &
        \begin{minipage}{0.180\textwidth}
            \includegraphics[width=0.462\textwidth, height=0.28\textwidth,cfbox=blue 1pt 0pt]{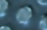}
            \includegraphics[width=0.462\textwidth, height=0.28\textwidth,cfbox=red 1pt 0pt]{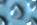}
        \end{minipage} &
        \begin{minipage}{0.180\textwidth}
            \includegraphics[width=0.462\textwidth, height=0.28\textwidth,cfbox=blue 1pt 0pt]{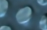}
            \includegraphics[width=0.462\textwidth, height=0.28\textwidth,cfbox=red 1pt 0pt]{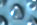}
        \end{minipage} &
        \begin{minipage}{0.180\textwidth}
            \includegraphics[width=0.462\textwidth, height=0.28\textwidth,cfbox=blue 1pt 0pt]{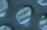}
            \includegraphics[width=0.462\textwidth, height=0.28\textwidth,cfbox=red 1pt 0pt]{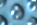}
        \end{minipage} \\
        \vspace{-8pt} \\
        % EPI
        &
        \begin{minipage}{0.180\textwidth}
            \includegraphics[width=0.462\textwidth, height=0.08\textwidth, cfbox=blue 1pt 0pt]{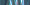}
            \includegraphics[width=0.462\textwidth, height=0.08\textwidth, cfbox=red 1pt 0pt]{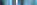}
        \end{minipage} &
        \begin{minipage}{0.180\textwidth}
            \includegraphics[width=0.462\textwidth, height=0.08\textwidth, cfbox=blue 1pt 0pt]{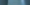}
            \includegraphics[width=0.462\textwidth, height=0.08\textwidth, cfbox=red 1pt 0pt]{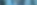}
        \end{minipage} &
        \begin{minipage}{0.180\textwidth}
            \includegraphics[width=0.462\textwidth, height=0.08\textwidth, cfbox=blue 1pt 0pt]{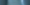}
            \includegraphics[width=0.462\textwidth, height=0.08\textwidth, cfbox=red 1pt 0pt]{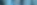}
        \end{minipage} &
        \begin{minipage}{0.180\textwidth}
            \includegraphics[width=0.462\textwidth, height=0.08\textwidth, cfbox=blue 1pt 0pt]{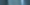}
            \includegraphics[width=0.462\textwidth, height=0.08\textwidth, cfbox=red 1pt 0pt]{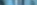}
        \end{minipage} &
        \begin{minipage}{0.180\textwidth}
            \includegraphics[width=0.462\textwidth, height=0.08\textwidth, cfbox=blue 1pt 0pt]{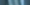}
            \includegraphics[width=0.462\textwidth, height=0.08\textwidth, cfbox=red 1pt 0pt]{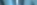}
        \end{minipage} \\
        \vspace{-8pt} \\
        % Two diffs
        &&
        \begin{minipage}{0.180\textwidth}
            \includegraphics[width=0.462\textwidth, height=0.28\textwidth,cfbox=blue 1pt 0pt]{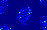}
            \includegraphics[width=0.462\textwidth, height=0.28\textwidth,cfbox=red 1pt 0pt]{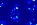}
        \end{minipage} &
        \begin{minipage}{0.180\textwidth}
            \includegraphics[width=0.462\textwidth, height=0.28\textwidth,cfbox=blue 1pt 0pt]{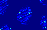}
            \includegraphics[width=0.462\textwidth, height=0.28\textwidth,cfbox=red 1pt 0pt]{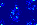}
        \end{minipage} &
        \begin{minipage}{0.180\textwidth}
            \includegraphics[width=0.462\textwidth, height=0.28\textwidth,cfbox=blue 1pt 0pt]{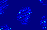}
            \includegraphics[width=0.462\textwidth, height=0.28\textwidth,cfbox=red 1pt 0pt]{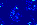}
        \end{minipage} &
        \begin{minipage}{0.180\textwidth}
            \includegraphics[width=0.462\textwidth, height=0.28\textwidth,cfbox=blue 1pt 0pt]{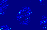}
            \includegraphics[width=0.462\textwidth, height=0.28\textwidth,cfbox=red 1pt 0pt]{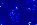}
        \end{minipage} \\
        \vspace{-7pt} \\
        % PSNR/SSIM
        \textit{Perforated\_Metal\_3} &&
        27.76/0.8966 &
        28.33/0.9077 &
        \underline{29.09}/\underline{0.9239} &
        \textbf{29.49}/\textbf{0.9312} \\
        &&&&& \green{+0.40}/\green{+0.0073} \\\hline
        \vspace{-3pt} \\

        % Building__Decoded
        % main image
        \multirow{4}{*}{\raisebox{10pt}[\height][0pt]{\includegraphics[width=0.180\textwidth, height=0.120\textwidth]{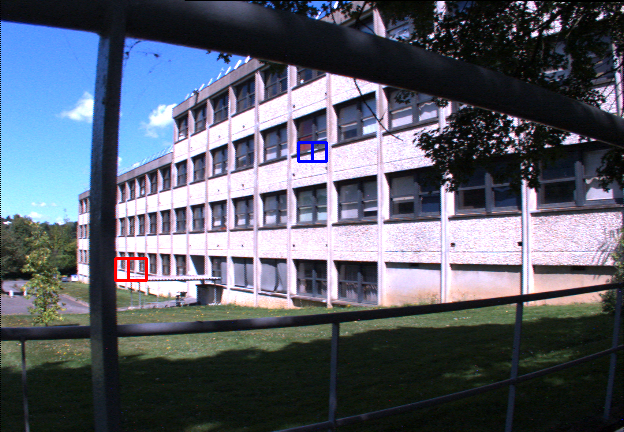}}} &
        % Two images
        \begin{minipage}{0.180\textwidth}
            \includegraphics[width=0.462\textwidth, height=0.28\textwidth,cfbox=blue 1pt 0pt]{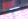}
            \includegraphics[width=0.462\textwidth, height=0.28\textwidth,cfbox=red 1pt 0pt]{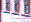}
        \end{minipage} &
        \begin{minipage}{0.180\textwidth}
            \includegraphics[width=0.462\textwidth, height=0.28\textwidth,cfbox=blue 1pt 0pt]{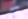}
            \includegraphics[width=0.462\textwidth, height=0.28\textwidth,cfbox=red 1pt 0pt]{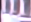}
        \end{minipage} &
        \begin{minipage}{0.180\textwidth}
            \includegraphics[width=0.462\textwidth, height=0.28\textwidth,cfbox=blue 1pt 0pt]{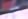}
            \includegraphics[width=0.462\textwidth, height=0.28\textwidth,cfbox=red 1pt 0pt]{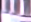}
        \end{minipage} &
        \begin{minipage}{0.180\textwidth}
            \includegraphics[width=0.462\textwidth, height=0.28\textwidth,cfbox=blue 1pt 0pt]{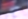}
            \includegraphics[width=0.462\textwidth, height=0.28\textwidth,cfbox=red 1pt 0pt]{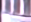}
        \end{minipage} &
        \begin{minipage}{0.180\textwidth}
            \includegraphics[width=0.462\textwidth, height=0.28\textwidth,cfbox=blue 1pt 0pt]{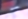}
            \includegraphics[width=0.462\textwidth, height=0.28\textwidth,cfbox=red 1pt 0pt]{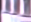}
        \end{minipage} \\
        \vspace{-8pt} \\
        % EPI
        &
        \begin{minipage}{0.180\textwidth}
            \includegraphics[width=0.462\textwidth, height=0.08\textwidth, cfbox=blue 1pt 0pt]{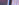}
            \includegraphics[width=0.462\textwidth, height=0.08\textwidth, cfbox=red 1pt 0pt]{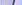}
        \end{minipage} &
        \begin{minipage}{0.180\textwidth}
            \includegraphics[width=0.462\textwidth, height=0.08\textwidth, cfbox=blue 1pt 0pt]{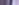}
            \includegraphics[width=0.462\textwidth, height=0.08\textwidth, cfbox=red 1pt 0pt]{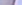}
        \end{minipage} &
        \begin{minipage}{0.180\textwidth}
            \includegraphics[width=0.462\textwidth, height=0.08\textwidth, cfbox=blue 1pt 0pt]{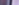}
            \includegraphics[width=0.462\textwidth, height=0.08\textwidth, cfbox=red 1pt 0pt]{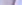}
        \end{minipage} &
        \begin{minipage}{0.180\textwidth}
            \includegraphics[width=0.462\textwidth, height=0.08\textwidth, cfbox=blue 1pt 0pt]{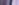}
            \includegraphics[width=0.462\textwidth, height=0.08\textwidth, cfbox=red 1pt 0pt]{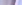}
        \end{minipage} &
        \begin{minipage}{0.180\textwidth}
            \includegraphics[width=0.462\textwidth, height=0.08\textwidth, cfbox=blue 1pt 0pt]{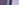}
            \includegraphics[width=0.462\textwidth, height=0.08\textwidth, cfbox=red 1pt 0pt]{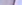}
        \end{minipage} \\
        \vspace{-8pt} \\
        % Two diffs
        &&
        \begin{minipage}{0.180\textwidth}
            \includegraphics[width=0.462\textwidth, height=0.28\textwidth,cfbox=blue 1pt 0pt]{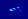}
            \includegraphics[width=0.462\textwidth, height=0.28\textwidth,cfbox=red 1pt 0pt]{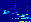}
        \end{minipage} &
        \begin{minipage}{0.180\textwidth}
            \includegraphics[width=0.462\textwidth, height=0.28\textwidth,cfbox=blue 1pt 0pt]{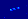}
            \includegraphics[width=0.462\textwidth, height=0.28\textwidth,cfbox=red 1pt 0pt]{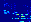}
        \end{minipage} &
        \begin{minipage}{0.180\textwidth}
            \includegraphics[width=0.462\textwidth, height=0.28\textwidth,cfbox=blue 1pt 0pt]{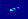}
            \includegraphics[width=0.462\textwidth, height=0.28\textwidth,cfbox=red 1pt 0pt]{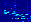}
        \end{minipage} &
        \begin{minipage}{0.180\textwidth}
            \includegraphics[width=0.462\textwidth, height=0.28\textwidth,cfbox=blue 1pt 0pt]{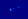}
            \includegraphics[width=0.462\textwidth, height=0.28\textwidth,cfbox=red 1pt 0pt]{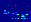}
        \end{minipage} \\
        \vspace{-7pt} \\
        % PSNR/SSIM
        \textit{Building\_Decoded} &&
        29.54/0.9309 &
        29.51/0.9319 &
        \underline{29.83}/\underline{0.9361} &
        \textbf{30.06}/\textbf{0.9383} \\
        &&&&& \green{+0.23}/\green{+0.0022} \\\hline
        \vspace{-3pt} \\

        % Lego Knights
        % main image
        \multirow{4}{*}{\raisebox{10pt}[\height][0pt]{\includegraphics[width=0.180\textwidth, height=0.130\textwidth]{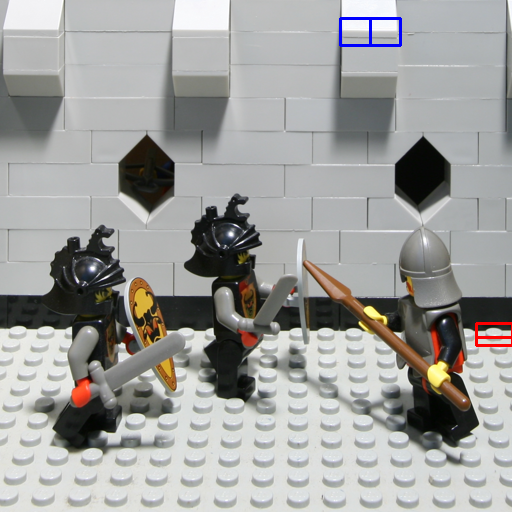}}} &
        % Two images
        \begin{minipage}{0.180\textwidth}
            \includegraphics[width=0.462\textwidth, height=0.28\textwidth,cfbox=blue 1pt 0pt]{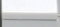}
            \includegraphics[width=0.462\textwidth, height=0.28\textwidth,cfbox=red 1pt 0pt]{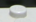}
        \end{minipage} &
        \begin{minipage}{0.180\textwidth}
            \includegraphics[width=0.462\textwidth, height=0.28\textwidth,cfbox=blue 1pt 0pt]{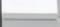}
            \includegraphics[width=0.462\textwidth, height=0.28\textwidth,cfbox=red 1pt 0pt]{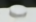}
        \end{minipage} &
        \begin{minipage}{0.180\textwidth}
            \includegraphics[width=0.462\textwidth, height=0.28\textwidth,cfbox=blue 1pt 0pt]{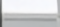}
            \includegraphics[width=0.462\textwidth, height=0.28\textwidth,cfbox=red 1pt 0pt]{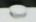}
        \end{minipage} &
        \begin{minipage}{0.180\textwidth}
            \includegraphics[width=0.462\textwidth, height=0.28\textwidth,cfbox=blue 1pt 0pt]{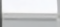}
            \includegraphics[width=0.462\textwidth, height=0.28\textwidth,cfbox=red 1pt 0pt]{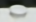}
        \end{minipage} &
        \begin{minipage}{0.180\textwidth}
            \includegraphics[width=0.462\textwidth, height=0.28\textwidth,cfbox=blue 1pt 0pt]{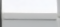}
            \includegraphics[width=0.462\textwidth, height=0.28\textwidth,cfbox=red 1pt 0pt]{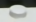}
        \end{minipage} \\
        \vspace{-8pt} \\
        % EPI
        &
        \begin{minipage}{0.180\textwidth}
            \includegraphics[width=0.462\textwidth, height=0.08\textwidth, cfbox=blue 1pt 0pt]{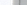}
            \includegraphics[width=0.462\textwidth, height=0.08\textwidth, cfbox=red 1pt 0pt]{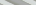}
        \end{minipage} &
        \begin{minipage}{0.180\textwidth}
            \includegraphics[width=0.462\textwidth, height=0.08\textwidth, cfbox=blue 1pt 0pt]{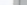}
            \includegraphics[width=0.462\textwidth, height=0.08\textwidth, cfbox=red 1pt 0pt]{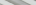}
        \end{minipage} &
        \begin{minipage}{0.180\textwidth}
            \includegraphics[width=0.462\textwidth, height=0.08\textwidth, cfbox=blue 1pt 0pt]{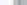}
            \includegraphics[width=0.462\textwidth, height=0.08\textwidth, cfbox=red 1pt 0pt]{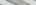}
        \end{minipage} &
        \begin{minipage}{0.180\textwidth}
            \includegraphics[width=0.462\textwidth, height=0.08\textwidth, cfbox=blue 1pt 0pt]{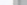}
            \includegraphics[width=0.462\textwidth, height=0.08\textwidth, cfbox=red 1pt 0pt]{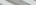}
        \end{minipage} &
        \begin{minipage}{0.180\textwidth}
            \includegraphics[width=0.462\textwidth, height=0.08\textwidth, cfbox=blue 1pt 0pt]{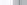}
            \includegraphics[width=0.462\textwidth, height=0.08\textwidth, cfbox=red 1pt 0pt]{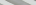}
        \end{minipage} \\
        \vspace{-8pt} \\
        % Two diffs
        &&
        \begin{minipage}{0.180\textwidth}
            \includegraphics[width=0.462\textwidth, height=0.28\textwidth,cfbox=blue 1pt 0pt]{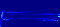}
            \includegraphics[width=0.462\textwidth, height=0.28\textwidth,cfbox=red 1pt 0pt]{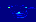}
        \end{minipage} &
        \begin{minipage}{0.180\textwidth}
            \includegraphics[width=0.462\textwidth, height=0.28\textwidth,cfbox=blue 1pt 0pt]{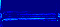}
            \includegraphics[width=0.462\textwidth, height=0.28\textwidth,cfbox=red 1pt 0pt]{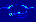}
        \end{minipage} &
        \begin{minipage}{0.180\textwidth}
            \includegraphics[width=0.462\textwidth, height=0.28\textwidth,cfbox=blue 1pt 0pt]{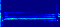}
            \includegraphics[width=0.462\textwidth, height=0.28\textwidth,cfbox=red 1pt 0pt]{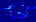}
        \end{minipage} &
        \begin{minipage}{0.180\textwidth}
            \includegraphics[width=0.462\textwidth, height=0.28\textwidth,cfbox=blue 1pt 0pt]{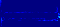}
            \includegraphics[width=0.462\textwidth, height=0.28\textwidth,cfbox=red 1pt 0pt]{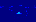}
        \end{minipage} \\
        \vspace{-7pt} \\
        % PSNR/SSIM
        \textit{Lego Knights} &&
        \underline{35.71}/\underline{0.9813} &
        35.44/0.9799 &
        35.38/0.9808 &
        \textbf{36.83}/\textbf{0.9850} \\
        &&&&& \green{+1.12}/\green{+0.0037} \\
    \end{tabular}
    }
    \caption{Qualitative comparison of $4\times$ LFSR on samples. For each sample, two regions are highlighted by blue and red bounding boxes. The zoom-in views, EPI, and error maps of these two regions are presented. PSNR/SSIM results of the whole LF image are given at the bottom.}
    \label{fig:Qual}
    \vspace{-5pt}
\end{figure*}

%% file: tex/figures/psnr2efficiency.tex
\begin{figure*}[!t]
    \centering
    \tabcolsep=0.06cm
    \begin{tabular}{ccc}
        (a) PSNR vs. Parameter Number &
        (b) PSNR vs. FLOPs &
        (c) PSNR vs. Inference time \\
        \includegraphics[width=0.28\textwidth, height=0.20\textwidth]{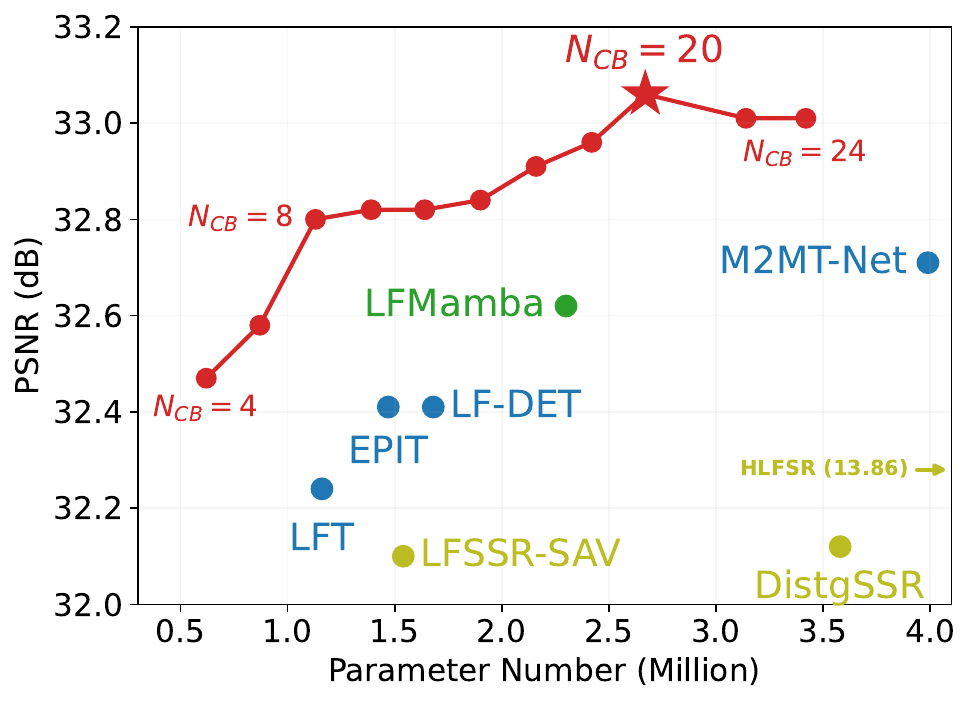} &
        \includegraphics[width=0.28\textwidth, height=0.20\textwidth]{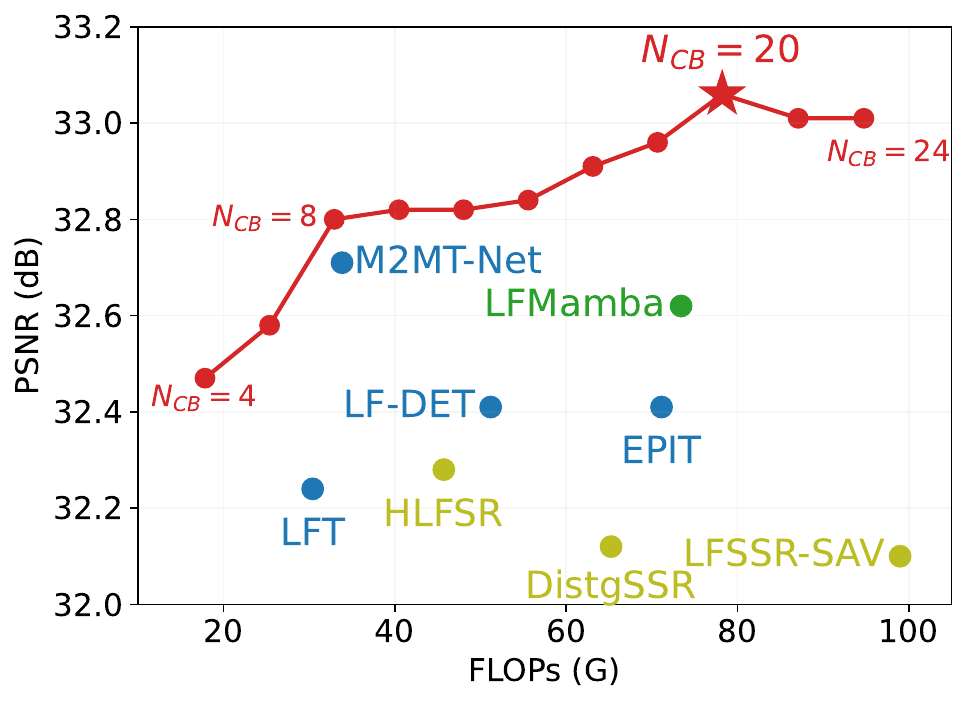} &
        \includegraphics[width=0.28\textwidth, height=0.20\textwidth]{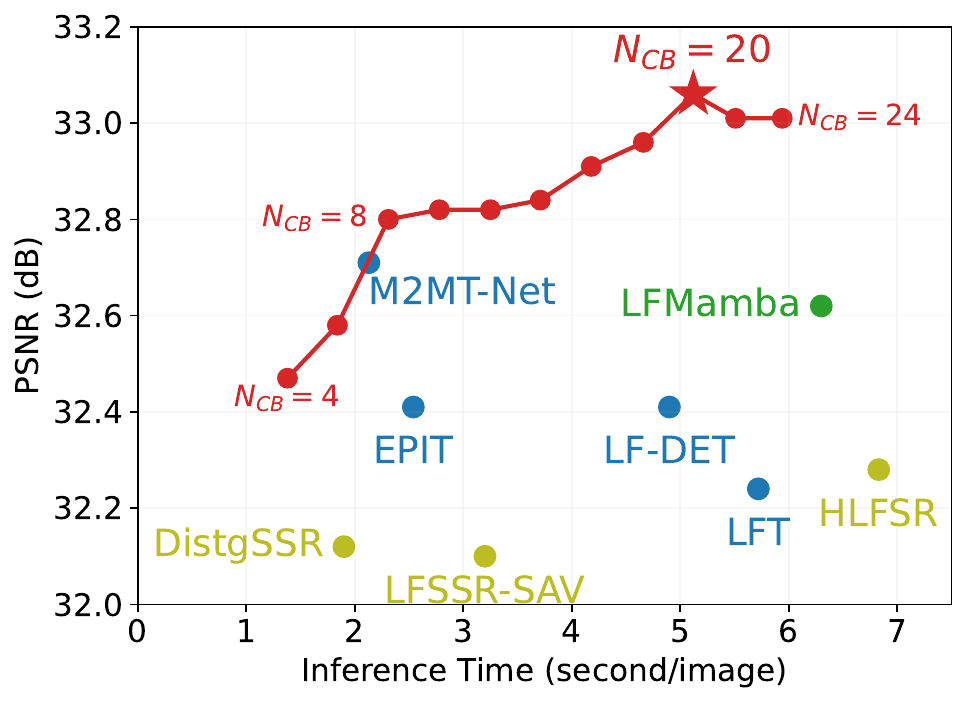} \\
        \vspace{-12pt}\\
        \multicolumn{3}{c}{\includegraphics[width=0.55\textwidth]{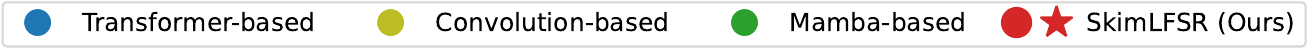}}
    \end{tabular}
    \caption{Trade-off between PSNR (the vertical axis) and model efficiency (the horizontal axis). Transformer-based methods are marked with blue dots, CNN-based methods are marked with yellow dots. The series of SkimLFSR variants are marked with red dots and connected with lines. Methods closer to the top right corner indicate better performance and efficiency.}
    \label{fig:psnr2efficiency_all}
    \vspace{-5pt}
\end{figure*}

%% file: tex/figures/dsa_features.tex
\begin{figure}[t!]
    \centering
    \tabcolsep=0.02cm
    \renewcommand{\arraystretch}{0.8}
    \resizebox{0.46\textwidth}{!}{
    \begin{tabular}{cccc}
        % Features
        \raisebox{1.4\height}{
            \begin{tabular}{c}
                \includegraphics[width=0.06\textwidth, height=0.06\textwidth]{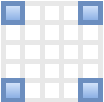} \\
                $DSA_1$
            \end{tabular}
        } &
        \includegraphics[width=0.200\textwidth, height=0.150\textwidth]{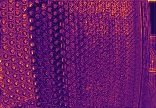} &
        \includegraphics[width=0.200\textwidth, height=0.150\textwidth]{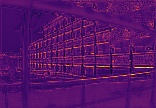} &
        \includegraphics[width=0.200\textwidth, height=0.150\textwidth]{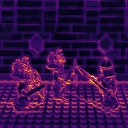} \\
        \raisebox{1.4\height}{
            \begin{tabular}{c}
                \includegraphics[width=0.06\textwidth, height=0.06\textwidth]{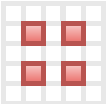} \\
                $DSA_2$
            \end{tabular}
        } &
        \includegraphics[width=0.200\textwidth, height=0.150\textwidth]{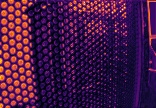} &
        \includegraphics[width=0.200\textwidth, height=0.150\textwidth]{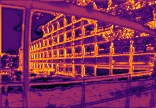} &
        \includegraphics[width=0.200\textwidth, height=0.150\textwidth]{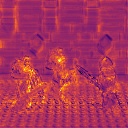} \\
        
        \raisebox{4.5\height}{$DSA_1 - DSA_2$} &
        \includegraphics[width=0.200\textwidth, height=0.150\textwidth]{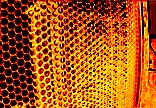} &
        \includegraphics[width=0.200\textwidth, height=0.150\textwidth]{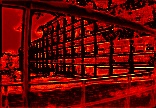} &
        \includegraphics[width=0.200\textwidth, height=0.150\textwidth]{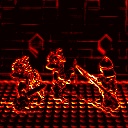} \\
        
        \raisebox{4.5\height}{$DSA_2 - DSA_1$} &
        \includegraphics[width=0.200\textwidth, height=0.150\textwidth]{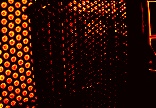} &
        \includegraphics[width=0.200\textwidth, height=0.150\textwidth]{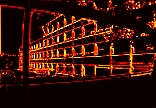} &
        \includegraphics[width=0.200\textwidth, height=0.150\textwidth]{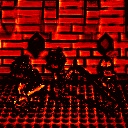} \\
        % Title
        &
        (a) \textit{Perforated\_Metal\_3} &
        (b) \textit{Building\_Decoded} &
        (c) \textit{Lego Knights} \\
    \end{tabular}
    }
    \caption{Visualization of DSA feature maps. The first two columns display the two DSA branches' output. The third and fourth columns present their differential activations, revealing the exclusive attention of each branch.}
    \label{fig:DSA_features}
    % \vspace{-10pt}
\end{figure}

%% file: tex/figures/DDR.tex
\begin{figure*}[tbp]
\begin{center}
    \renewcommand{\arraystretch}{0.9}
    \tabcolsep=0.01cm
    \resizebox{0.90\textwidth}{!}{
    \begin{tabular}{cccc}
        (a) EPIT & (b) LF-DET & (c) M2MT-Net & (d) SkimLFSR (Ours) \\
        CHI: 17.25 & CHI: 6.48 & CHI: 11.86 & \textbf{CHI: 41.27} \\
        \includegraphics[width=0.25\textwidth, height=0.19\textwidth]{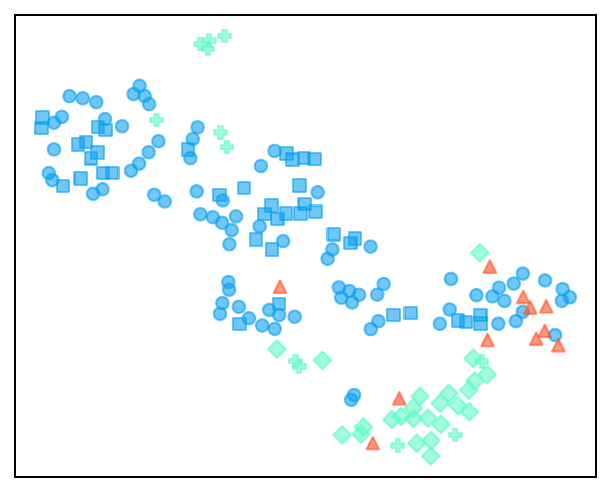} &
        \includegraphics[width=0.25\textwidth, height=0.19\textwidth]{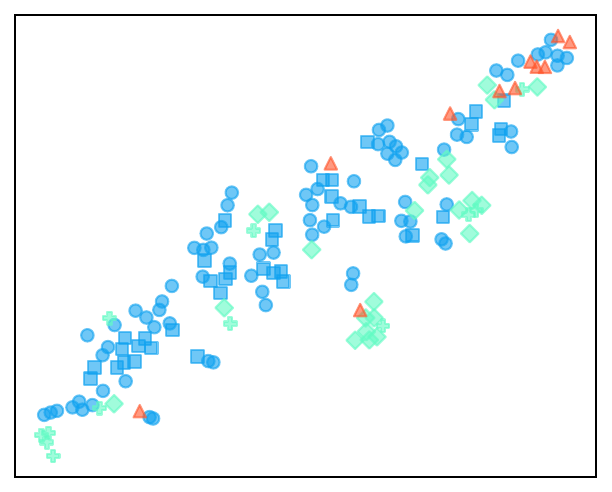} &
        \includegraphics[width=0.25\textwidth, height=0.19\textwidth]{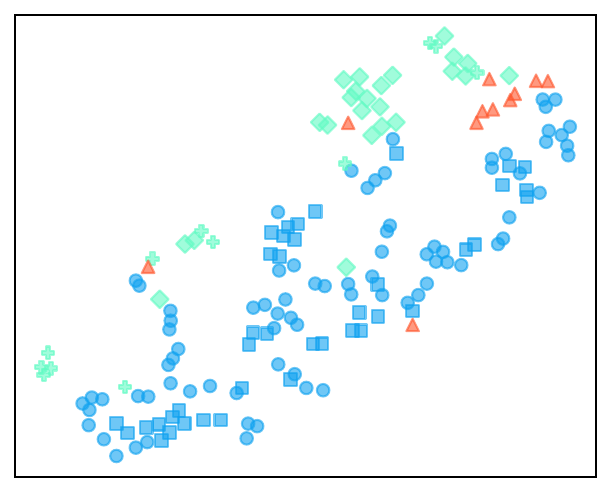} &
        \includegraphics[width=0.25\textwidth, height=0.19\textwidth]{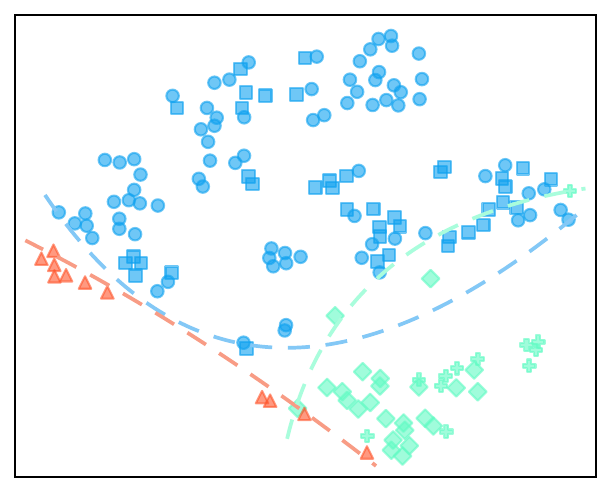} \\
        \vspace{-5pt}\\
        \multicolumn{4}{c}{
            \hspace{-12pt}\includegraphics[width=0.70\textwidth]{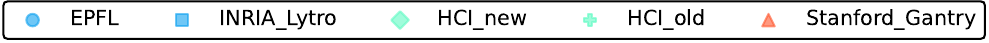}
        }
    \end{tabular}
    }
    \caption{Projected feature representation using t-SNE. Each plot contains 167 points, each of which represents a sample in the dataset.}
    \label{fig:DDR}
\end{center}
\end{figure*}

%% file: tex/tables/dsa_variants.tex
\begin{table*}[t!]
    \caption{Comparison of different skimmed SAI sets for SkimLFSR ($N_{CB}=6$) in $4\times$ LFSR.}
    \label{tab:DSA_variants}
    \centering
    \tabcolsep=0.05cm
    \resizebox{0.90\textwidth}{!}{
    \begin{tabular}{|l|c|c|c|c|c|c|c|c|c|}
        \hline
        \multirow{3}{*}[-4pt]{Variants} &
        (a) & (b) & (c) & (d) & (e) & (f) & (g) & (h) & (i) \\
        &
        \includegraphics[width=0.035\textwidth, height=0.035\textwidth, valign=m]{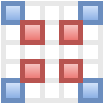} &
        \includegraphics[width=0.035\textwidth, height=0.035\textwidth, valign=m]{images/dsa_indicators/DSA_1.pdf} &
        \includegraphics[width=0.035\textwidth, height=0.035\textwidth, valign=m]{images/dsa_indicators/DSA_2.pdf} &
        \includegraphics[width=0.035\textwidth, height=0.035\textwidth, valign=m]{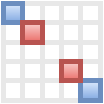} &
        \includegraphics[width=0.035\textwidth, height=0.035\textwidth, valign=m]{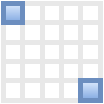} &
        \includegraphics[width=0.035\textwidth, height=0.035\textwidth, valign=m]{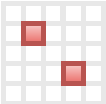} &
        \includegraphics[width=0.035\textwidth, height=0.035\textwidth, valign=m]{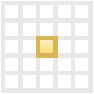} &
        \includegraphics[width=0.035\textwidth, height=0.035\textwidth, valign=m]{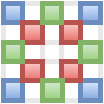} &
        \includegraphics[width=0.035\textwidth, height=0.035\textwidth, valign=m]{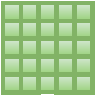} \\
        &&&&&&&&& \vspace*{-6pt}\\\hline
        PSNR/SSIM           & 32.58/0.9474  & 32.47/0.9451  & 32.45/0.9450  & 32.40/0.9449  & 32.30/0.9438  & 32.30/0.9435  & 32.05/0.9425  & 32.57/0.9469 & 32.50/0.9450  \\\hline
        \#Branches          & 2             & 1             & 1             & 2             & 1             & 1             & 1             & 3            & 1             \\\hline
        \#SAIs              & 4+4=8         & 4             & 4             & 2+2=4         & 2             & 2             & 1             & 4+4+4=12     & 25            \\\hline
        \#Parameters (M)    & 0.87          & 0.82          & 0.82          & 0.82          & 0.76          & 0.76          & 0.73          & 0.93         & 1.40          \\\hline
        FLOPs (G)           & 25.38         & 25.03         & 25.03         & 25.33         & 24.97         & 24.97         & 24.94         & 25.75        & 25.62         \\\hline
        Inference Time (s)  & 1.79          & 1.76          & 1.76          & 1.79          & 1.73          & 1.74          & 1.74          & 1.88         & 1.77          \\\hline
    \end{tabular}
    }
    \vspace{-5pt}
\end{table*}

%% file: tex/figures/dsa_12_vs_dsa_full.tex
\begin{figure}[t!]
    \centering
    \includegraphics[width=0.30\textwidth]{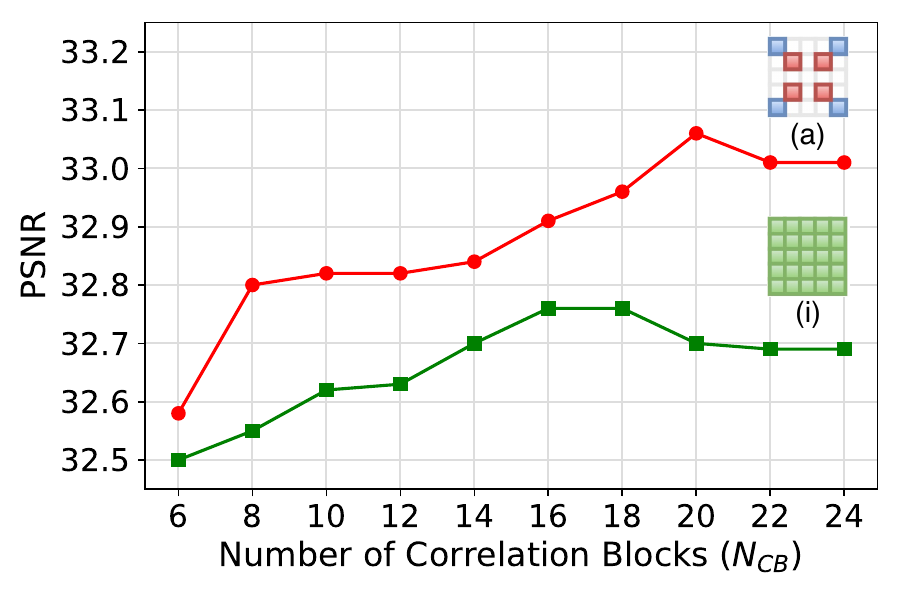}
    \vspace{-7pt}
    \caption{Comparison of performance of the double-branch baseline and the single-branch variant (Variant (a) and Variant (i) in Table \ref{tab:DSA_variants}) varying the number of correlation blocks $N_{CB}$.}
    \label{fig:dsa_12_vs_dsa_full}
    \vspace{-10pt}
\end{figure}

%% file: tex/tables/OverallComparison_7x7.tex
\begin{table*}[ht!]
\caption{\hl{Comparison of PSNR/SSIM on $7 \times 7$ SAIs and $4 \times$ LFSR. In the first group, the best and second-best results are highlighted in bold and underlined, respectively. In the second and third groups, the difference between the SkimLFSR baseline model and its variants is displayed below PSNR/SSIM.}}
\centering
\resizebox{0.95\textwidth}{!}{
\begin{threeparttable}

\begin{tabular}{|l|c|c|c|c|c|c|}
    \hline
    \makecell[l]{Methods\\(\#Train/\#Test samples)} & \makecell{\textit{EPFL}\\(70/10)} & \makecell{\textit{HCInew}\\(20/4)} & \makecell{\textit{HCIold}\\(10/2)} & \makecell{\textit{INRIA}\\(35/5)} & \makecell{\textit{STFgantry}\\(9/2)} & Average \\    \hline\hline
    \multicolumn{7}{|l|}{\textbf{1. Qualitative Comparison (Trained on $7 \times 7$ SAIs)}} \\\hline
    LF-DET \cite{congLFDET_TMM2024}	            & 30.07/\underline{0.9314}                  & \underline{31.94}/\underline{0.9286}  & 38.10/0.9756                              & 31.93/0.9578                              & 32.57/0.9604                            & 32.92/0.9508                            \\
    M2MT-Net \cite{hu_M2MTNet_TMM2024}	        & \underline{30.14}/\underline{0.9314}      & 31.80/0.9280                          & \underline{38.22}/\underline{0.9761}      & \underline{31.99}/\underline{0.9579}      & \underline{32.68}/\underline{0.9606}    & \underline{32.97}/\underline{0.9508}    \\
    \hline
    \rule[-4pt]{0pt}{12pt}\multirow{2}{*}{SkimLFSR (Ours) \adjustbox{valign=m}{\includegraphics[width=0.035\textwidth]{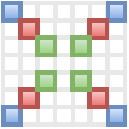}}}
    & \textbf{30.40}/\textbf{0.9365}            & \textbf{32.06}/\textbf{0.9304}        & \textbf{38.33}/\textbf{0.9765}            & \textbf{32.11}/\textbf{0.9601}            & \textbf{32.96}/\textbf{0.9631}          & \textbf{33.15}/\textbf{0.9532}          \\
    & \green{+0.26}/\green{+0.0051}             & \green{+0.12}/\green{+0.0018}         & \green{+0.11}/\green{+0.0004}             & \green{+0.12}/\green{+0.0022}             & \green{+0.28}/\green{+0.0025}           & \green{+0.18}/\green{+0.0024}           \\
    \hline\hline
    \multicolumn{7}{|l|}{\textbf{2. Ablation Study on Skimmed SAI Sets (Trained on $7 \times 7$ SAIs)}} \\\hline
    \rule[-4pt]{0pt}{12pt}\multirow{2}{*}{(a) \adjustbox{valign=m}{\includegraphics[width=0.035\textwidth]{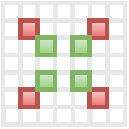}}}
                                                & 30.30/0.9357                              & 32.04/0.9300                          & 38.26/0.9763                              & 31.96/0.9596                              & 32.81/0.9625                            & 33.07/0.9528                    \\
                                                & \red{-0.10}/\red{-0.0008}                 & \red{-0.02}/\red{-0.0004}             & \red{-0.07}/\red{-0.0002}                 & \red{-0.15}/\red{-0.0005}                 & \red{-0.15}/\red{-0.0006}               & \red{-0.08}/\red{-0.0004}       \\\hline
    \rule[-4pt]{0pt}{12pt}\multirow{2}{*}{(b) \adjustbox{valign=m}{\includegraphics[width=0.035\textwidth]{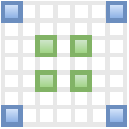}}}
                                                & 30.26/0.9352                              & 32.03/0.9299                          & 38.25/0.9762                              & 31.94/0.9595                              & 32.85/0.9628                            & 33.07/0.9528                    \\
                                                & \red{-0.14}/\red{-0.0013}                 & \red{-0.03}/\red{-0.0005}             & \red{-0.08}/\red{-0.0003}                 & \red{-0.17}/\red{-0.0006}                 & \red{-0.11}/\red{-0.0003}               & \red{-0.08}/\red{-0.0004}       \\\hline
    \rule[-4pt]{0pt}{12pt}\multirow{2}{*}{(c) \adjustbox{valign=m}{\includegraphics[width=0.035\textwidth]{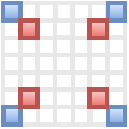}}}
                                                & 30.22/0.9354                              & 32.02/0.9298                          & 38.24/0.9763                              & 31.92/0.9594                              & 32.81/0.9626                            & 33.04/0.9527                    \\
                                                & \red{-0.18}/\red{-0.0011}                 & \red{-0.04}/\red{-0.0006}             & \red{-0.09}/\red{-0.0002}                 & \red{-0.19}/\red{-0.0007}                 & \red{-0.15}/\red{-0.0005}               & \red{-0.11}/\red{-0.0005}       \\
    \hline\hline
    \multicolumn{7}{|l|}{\textbf{3. Cross-angular-resolution Generalization (Trained on $5 \times 5$ SAIs $\rightarrow$ and tested on $7 \times 7$ SAIs)}} \\\hline
    DistgSSR \cite{wangDistgSSR_TPAMI2023}	    & N/A                                       & N/A                                   & N/A                                       & N/A                                       & N/A                                     & N/A                             \\
    HLFSR \cite{duongHLFSR_TCI2023}	            & N/A                                       & N/A                                   & N/A                                       & N/A                                       & N/A                                     & N/A                             \\
    LFT \cite{liangLFT_SPL2022}	                & 28.67/0.9112                              & 30.15/0.8989                          & 36.22/0.9636                              & 30.66/0.9447                              & 29.54/0.9220                            & 31.05/0.9281                    \\
    EPIT \cite{liangEPIT_ICCV2023}	            & 29.58/0.9230                              & 31.63/0.9250                          & 37.93/0.9749                              & 31.49/0.9538                              & 32.40/0.9585                            & 32.61/0.9470                    \\
    LF-DET \cite{congLFDET_TMM2024}	            & 29.68/0.9250                              & 31.58/0.9239                          & 37.90/0.9745                              & 31.51/0.9541                              & 32.14/0.9567                            & 32.56/0.9469                    \\
    LFMamba \cite{xia_LFMamba_arxiv2024}	    & 29.76/0.9270                              & 31.34/0.9237                          & 37.18/0.9717                              & 31.63/0.9545                              & 31.55/0.9535                            & 32.29/0.9461                    \\
    M2MTNet \cite{hu_M2MTNet_TMM2024}	        & N/A                                       & N/A                                   & N/A                                       & N/A                                       & N/A                                     & N/A                             \\\hline
    \rule[-4pt]{0pt}{12pt}\multirow{2}{*}{(d) \adjustbox{valign=m}{\includegraphics[width=0.025\textwidth]{images/dsa_indicators/DSA_12.pdf}} $\rightarrow$ \adjustbox{valign=m}{\includegraphics[width=0.035\textwidth]{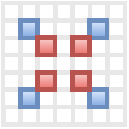}}}
                                                & 30.21/0.9328                              & 31.96/0.9284                          & 38.18/0.9757                              & 31.99/0.9581                              & 32.71/0.9610                            & 33.01/0.9512                    \\
                                                & \red{-0.19}/\red{-0.0037}                 & \red{-0.10}/\red{-0.0020}             & \red{-0.15}/\red{-0.0008}                 & \red{-0.12}/\red{-0.0020}                 & \red{-0.25}/\red{-0.0021}               & \red{-0.14}/\red{-0.0020}       \\\hline
    \rule[-4pt]{0pt}{12pt}\multirow{2}{*}{(e) \adjustbox{valign=m}{\includegraphics[width=0.025\textwidth]{images/dsa_indicators/DSA_12.pdf}} $\rightarrow$ \adjustbox{valign=m}{\includegraphics[width=0.035\textwidth]{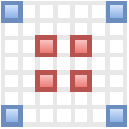}}}
                                                & 30.32/0.9337                              & 31.98/0.9284                          & 38.26/0.9762                              & 32.07/0.9584                              & 32.73/0.9610                            & 33.07/0.9515                    \\
                                                & \red{-0.08}/\red{-0.0028}                 & \red{-0.08}/\red{-0.0020}             & \red{-0.07}/\red{-0.0003}                 & \red{-0.04}/\red{-0.0017}                 & \red{-0.23}/\red{-0.0021}               & \red{-0.08}/\red{-0.0017}       \\\hline
    \rule[-4pt]{0pt}{12pt}\multirow{2}{*}{(f) \adjustbox{valign=m}{\includegraphics[width=0.025\textwidth]{images/dsa_indicators/DSA_12.pdf}} $\rightarrow$ \adjustbox{valign=m}{\includegraphics[width=0.035\textwidth]{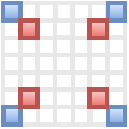}}}
                                                & 30.35/0.9338                              & 31.95/0.9273                          & 38.21/0.9758                              & 32.05/0.9582                              & 32.66/0.9602                            & 33.05/0.9511                    \\
                                                & \red{-0.05}/\red{-0.0027}                 & \red{-0.11}/\red{-0.0031}             & \red{-0.12}/\red{-0.0007}                 & \red{-0.06}/\red{-0.0019}                 & \red{-0.30}/\red{-0.0029}               & \red{-0.10}/\red{-0.0021}       \\\hline
\end{tabular}
\end{threeparttable}
}
\label{tab:overall_7x7}
\vspace{-3pt}
\end{table*}

%% file: tex/5.Conclusion.tex
\section{Conclusion and Future Work}
In this paper, we presented SkimLFSR, a novel and efficient LFSR network that overcomes the challenge of disparity entanglement in LF image processing. The experimental results validated SkimLFSR's \textbf{``less is more"} philosophy, demonstrating that SkimLFSR outperformed the state-of-the-art methods significantly by utilizing less but more pertinent information with reduced memory and computational complexity. Several analyses were conducted to reveal the underlying disparity-aware mechanisms and cross-angular-resolution generalizability of SkimLFSR.

Despite its success, our work has certain potential limitations that are worth further research.
\hl{First, the skimmed SAI subsets of the Skim Transformers are manually pre-defined. Although this is not a complex problem and is only needed once for each model, it essentially introduces a prior into the network, which could potentially limit the generalization ability when it comes to a new angular resolution or an irregular angular subspace. One potential solution is to make the process of SAI selection heuristic or learned in network optimization autonomously.}
\hl{Second, most LF images in the current BasicLFSR \cite{BasicLFSR} benchmark feature relatively simple objects and depth structures with limited intra-scene disparity variation. The light field community lacks systematic studies on how LFSR methods handle LF images with complex depth ranges within the same scene. Addressing this gap requires developing a dedicated dataset and evaluation protocol to systematically assess LFSR performance under complex intra-scene disparity variations in our future work.
}
\hl{Third, while LF cameras are typically designed with regular micro-lens layouts, resulting in uniform grid-structured angular sampling, extending SkimLFSR to handle non-uniform angular spaces remains an open challenge. The reliance on regular network operations, such as standard convolutions that assume uniform spatial relationships, presents a potential obstacle to this extension. An additional consideration is whether current LFSR methods can effectively handle the variable disparities that arise from non-uniform angular sampling across different SAIs within the same LF image. Addressing these challenges may require the development of adaptive architectures capable of processing irregularly sampled angular data.}

%% file: tex/Bio.tex
\begin{IEEEbiography}[{\includegraphics[width=1in,height=1.25in,clip,keepaspectratio]{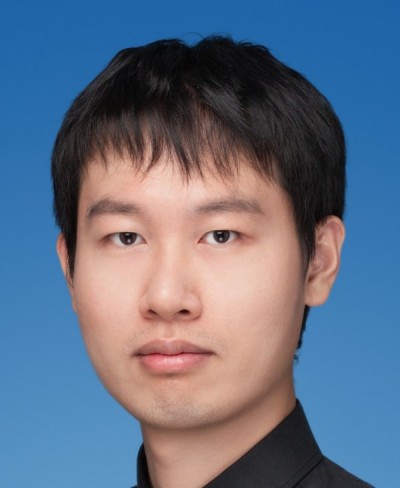}}]{Zeke Zexi Hu} received his Ph.D. degree from the School of Computer Science at the University of Sydney, Australia, in 2025, and his bachelor's degree in Software Engineering from South China Agricultural University, China, in 2014. His research focuses on image processing and visual tracking. He has authored and co-authored papers in academic journals and conferences, including IEEE TMM, TCSVT, TVCG, ICASSP, ICIP, etc.
\end{IEEEbiography}

\begin{IEEEbiography}[{\includegraphics[width=1in,height=1.25in,clip,keepaspectratio]{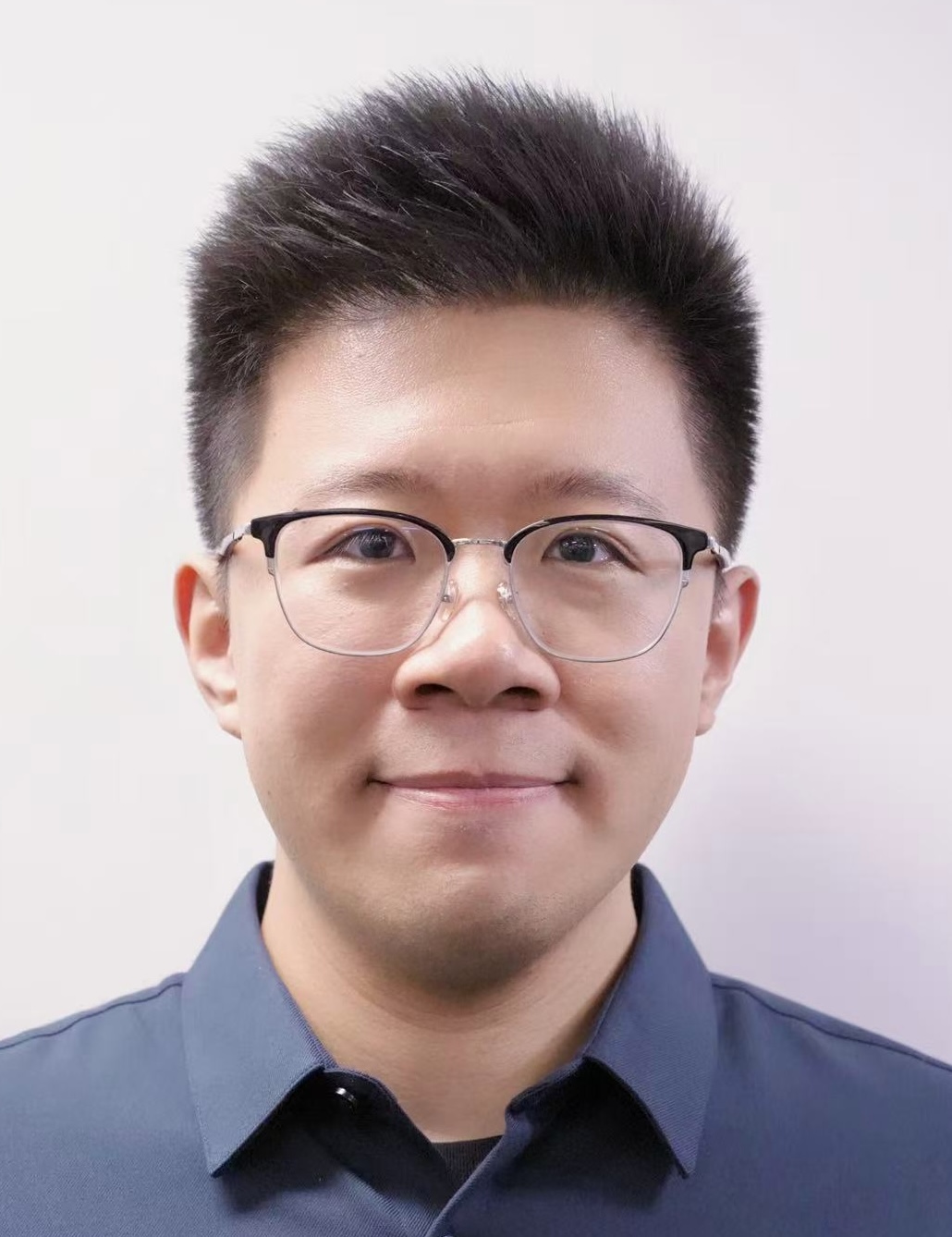}}]{Haodong Chen} received the Bachelor of Computer Science and Technology (Honours) from the University of Sydney, Australia, where he is currently pursuing the Ph.D. degree in Computer Science. His research focuses on applying artificial intelligence and bio-inspired sensors to computer vision and computer graphics, including image segmentation, edge detection, and 3D reconstruction.
\end{IEEEbiography}

\begin{IEEEbiography}[{\includegraphics[width=1in,height=1.25in,clip,keepaspectratio]{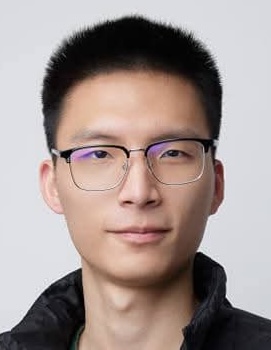}}]{Hui Ye} earned a Bachelor of Engineering from Chengdu University of Technology, China, and a Master of Engineering from Monash University, Australia. He later received a Master of Philosophy from the University of Sydney, where he is currently pursuing a Ph.D. in Computer Science. His research focuses on remote sensing, semantic segmentation, and vision-language models.
\end{IEEEbiography}

\begin{IEEEbiography}[{\includegraphics[width=1in,height=1.25in,clip,keepaspectratio]{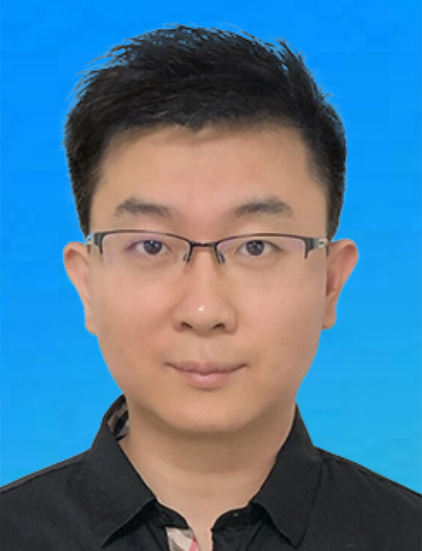}}]{Xiaoming Chen} holds a B.Sc. degree (with Distinction) from Royal Melbourne Institute of Technology and a Ph.D. degree from the University of Sydney, Australia. He has a combined experience in industry and academia. He has been with National University of Singapore, Nanyang Technological University, Singapore, CSIRO Australia, Technicolor Research, IBM Corporation, and University of Science and Technology of China (USTC). He is now a Professor at the School of Computer and Artificial Intelligence, Beijing Technology and Business University (BTBU), China, and a researcher at the University of Sydney, Australia. His research interests include immersive media computing, virtual reality, bio-inspired event processing, and related applications. His work has been published in journals and conferences including IEEE Trans. Vis. Comput. Graph., IEEE Trans. Image Process., IEEE Trans. Circuits Syst. Video Technol., IEEE Trans. Mult., IEEE VR, ACM MM, ECCV, AAAI, etc.
\end{IEEEbiography}

\begin{IEEEbiography}[{\includegraphics[width=1in,height=1.25in,clip,keepaspectratio]{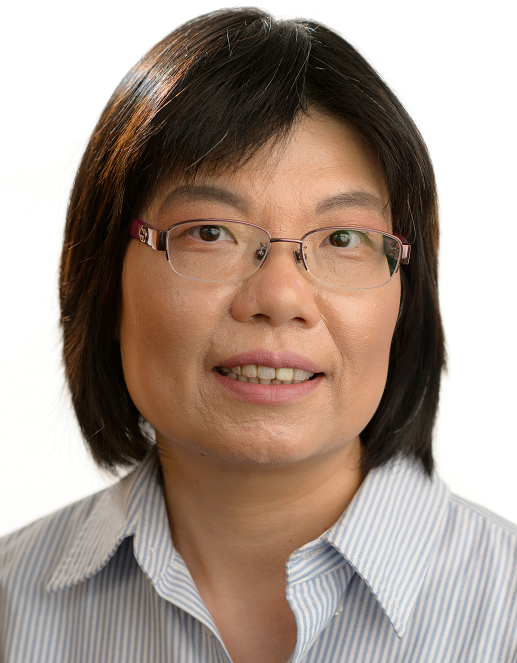}}]{Vera Yuk Ying Chung} received her B.S. degree in Computing and Information Systems from the University of London, UK, in 1995, and her Ph.D. degree in Computer Engineering from the Queensland University of Technology, Australia in 2000. She is a Senior Lecturer at the School of Computer Science, University of Sydney, Australia. Her research interests are in the areas of multimedia computing and virtual reality. Her work has been published in journals and conferences including IEEE Trans. on Image Processing, IEEE Trans. Vis. Comput. Graph., IEEE Trans. Mult., IEEE Trans. Circuits Syst. Video Technol., NeurIPS, ECCV, etc.
\end{IEEEbiography}

\begin{IEEEbiography}[{\includegraphics[width=1in,height=1.25in,clip,keepaspectratio]{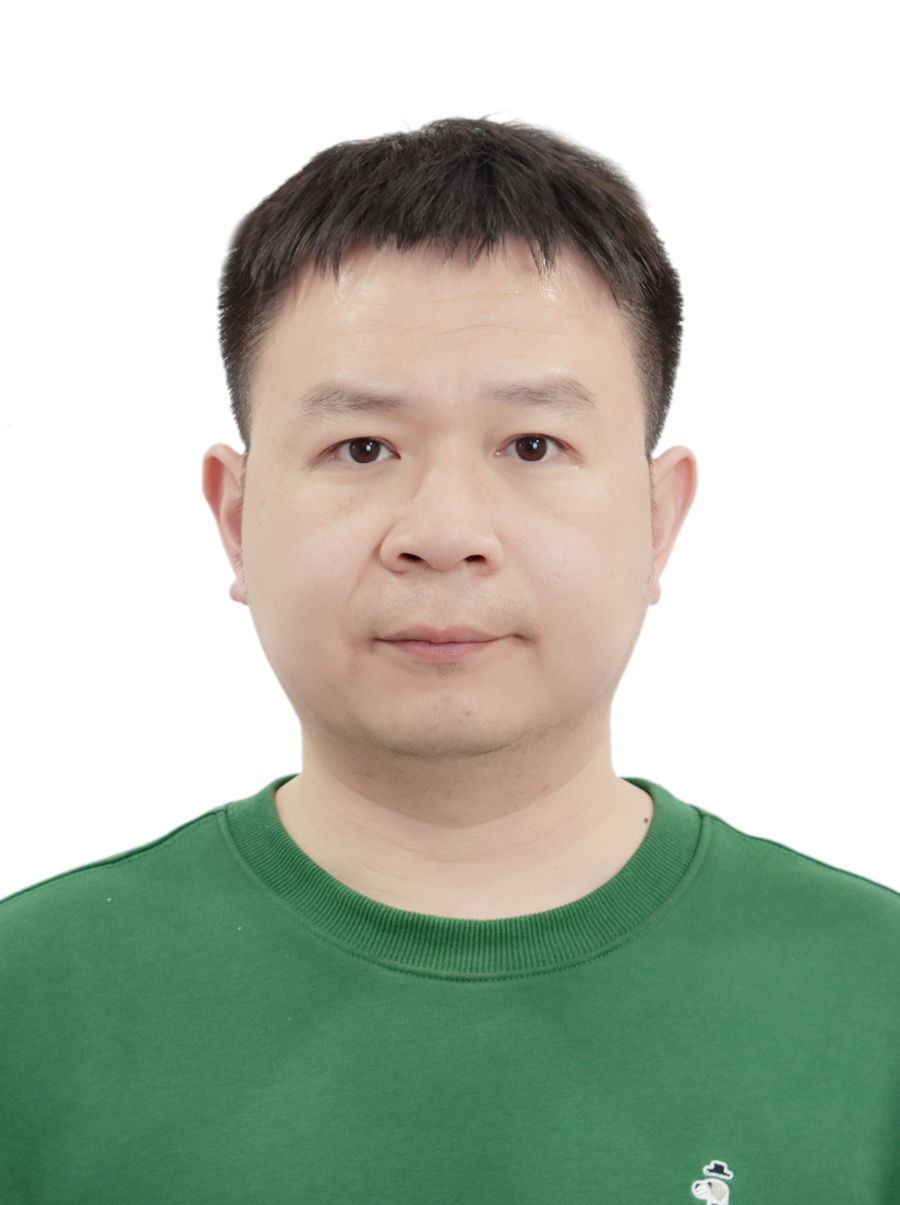}}]
{Yiran Shen} is professor in School of Software, Shandong University. He received his BE in communication engineering from Shandong University, China and his PhD degree in computer science and engineering from University of New South Wales. He published regularly at top-tier conferences and journals. Generally speaking, his research interest is sensing and computing for immersive systems. 
He is a Senior Member of IEEE. He has been regularly publishing papers in journals and conferences including IEEE Trans. on Pattern Analysis and Machine Intelligence, IEEE Trans. on Image Processing, IEEE Trans. on Visualization and Computer Graphics, IEEE Trans. on Mobile Computing, CVPR, NeurIPS, IEEE VR, etc. His work was awarded the Best Paper Award by IEEE VR 2024.

He is currently an Associate Editor of \textit{ACM Transactions on Sensor Networks} (\textit{TOSN}).
\end{IEEEbiography}

\begin{IEEEbiography}[{\includegraphics[width=1in,height=1.25in,clip,keepaspectratio]{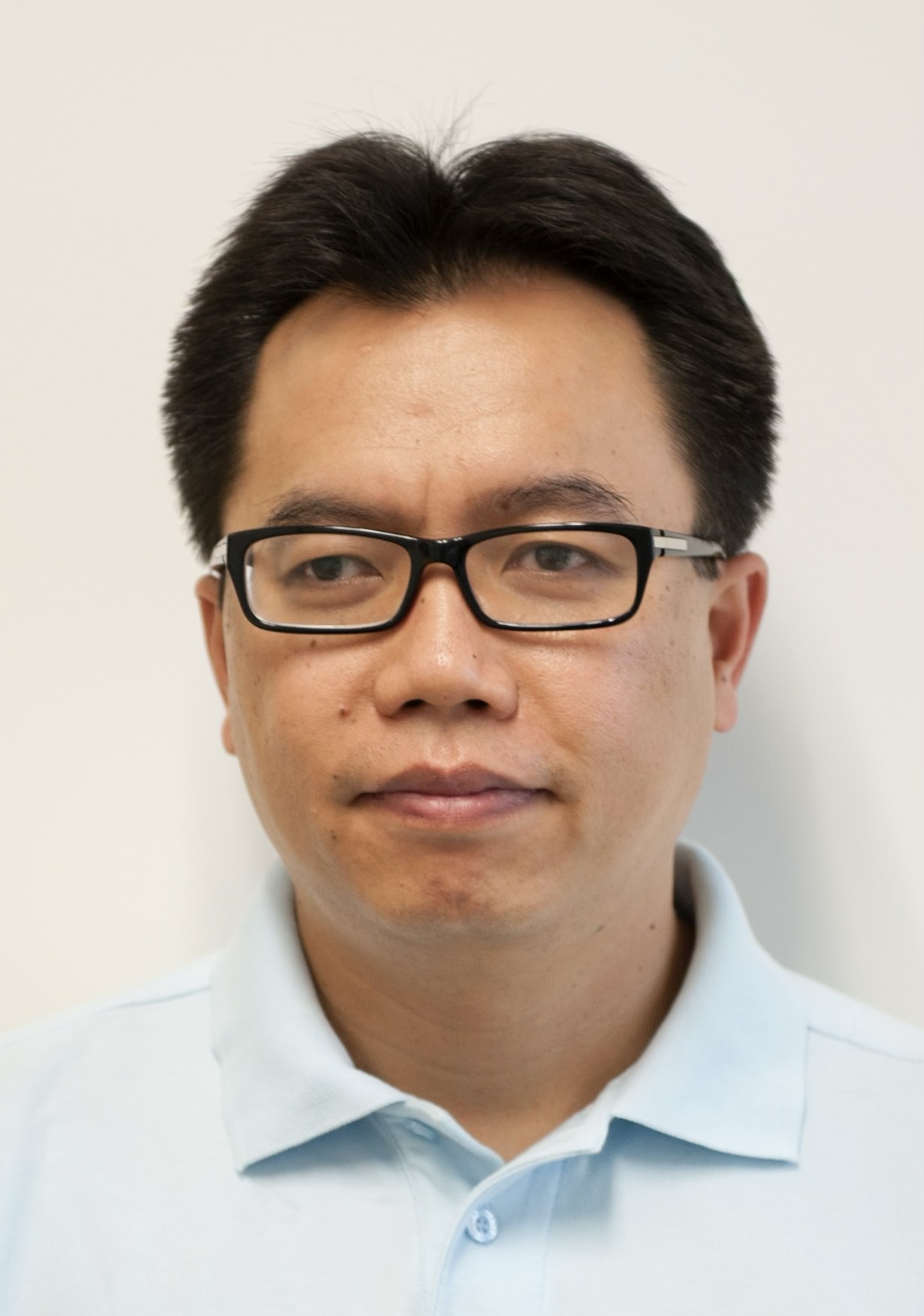}}]
{Weidong Cai} received the Ph.D. degree in computer science from the Basser Department of Computer Science, The University of Sydney, Australia, in 2001. He is currently an Associate Professor and the Director of Multimedia Laboratory, School of Computer Science, Faculty of Engineering, The University of Sydney. His research interests include multimedia computing, computer vision, medical image computing, machine learning, image/video processing, pattern recognition, and computer graphics. He has been regularly publishing papers in journals and conferences including IEEE Trans. on Image Processing, IEEE Trans. on Medical Imaging, IEEE Trans. on Visualization and Computer Graphics, International Journal of Computer Vision, CVPR, ICCV, ECCV, NeurIPS, etc.

He is currently a Senior Area Editor of \textit{IEEE Transactions on Image Processing} (\textit{TIP}).
\end{IEEEbiography}